\documentclass{article}
\usepackage[final]{neurips_2023}

\setcitestyle{authoryear,round,citesep={;},aysep={,},yysep={;}}

\usepackage[utf8]{inputenc} 
\usepackage[T1]{fontenc}    
\PassOptionsToPackage{hyphens}{url}
\usepackage{hyperref}
\usepackage{url}            
\usepackage{booktabs}       
\usepackage{amsfonts}       
\usepackage{nicefrac}       
\usepackage{microtype}      
\usepackage[table]{xcolor}  
\usepackage{placeins}       
\usepackage{ragged2e}
\usepackage{longtable}
\usepackage{doi}

\usepackage[most]{tcolorbox} 
\colorlet{lightgray}{gray!8}
\colorlet{darkgray}{gray!55}
\definecolor{myyellow}{HTML}{fff2cc}
\definecolor{myred}{HTML}{d72222}
\definecolor{myorange}{HTML}{F2C078}
\definecolor{mygreen}{HTML}{7EBC89}
\definecolor{myblue}{HTML}{0c234b}
\definecolor{myblack}{HTML}{071013}
\usetikzlibrary{positioning}

\definecolor{whitesmoke}{HTML}{F5F5F5}
\colorlet{background}{whitesmoke}

\newcolumntype{C}[1]{>{\centering\arraybackslash}p{#1}}%
\newcolumntype{D}[1]{>{\centering\arraybackslash}m{#1}}%

\usepackage[labelfont=bf]{caption}
\usepackage{changepage}

\usepackage{enumitem}

\usepackage[capitalize,noabbrev, nameinlink]{cleveref}

\title{Training Compute Thresholds:\\Features and Functions in AI Regulation}

\author{%
    \parbox{\linewidth}{%
        \centering\bfseries\large
        \begin{tabular}[t]{@{}c@{}}
            Lennart Heim\thanks{Correspondence to \href{mailto:lennart.heim@governance.ai}{\texttt{lennart.heim@governance.ai}}.} \\[5pt]
            \small\mdseries\begin{tabular}[c]{@{}c@{}}
                \rule{0pt}{1.5\normalbaselineskip} 
                Centre for the Governance of AI \\[0.4ex]
                Oxford, United Kingdom \\
                \rule{0pt}{1.5\normalbaselineskip} 
            \end{tabular}
        \end{tabular}%
        \hspace{3em}%
        \begin{tabular}[t]{@{}c@{}}
            Leonie Koessler \\[5pt]
            \small\mdseries\begin{tabular}[c]{@{}c@{}}
                Centre for the Governance of AI \\[0.4ex]
                Oxford, United Kingdom \& \\[0.7ex]
                European New School of Digital Studies \\[0.4ex]
                Frankfurt (Oder), Germany
            \end{tabular}
        \end{tabular}%
    }%
}

\begin{document}

\maketitle

\begin{abstract}
Regulators in the US and EU are using thresholds based on training compute—the number of computational operations used in training—to identify general-purpose artificial intelligence (GPAI) models that may pose risks of large-scale societal harm. We argue that training compute currently is the most suitable metric to identify GPAI models that deserve regulatory oversight and further scrutiny. Training compute correlates with model capabilities and risks, is quantifiable, can be measured early in the AI lifecycle, and can be verified by external actors, among other advantageous features. These features make compute thresholds considerably more suitable than other proposed metrics to serve as an initial filter to trigger additional regulatory requirements and scrutiny. However, training compute is an imperfect proxy for risk. As such, compute thresholds should not be used in isolation to determine appropriate mitigation measures. Instead, they should be used to detect potentially risky GPAI models that warrant regulatory oversight, such as through notification requirements, and further scrutiny, such as via model evaluations and risk assessments, the results of which may inform which mitigation measures are appropriate. In fact, this appears largely consistent with how compute thresholds are used today. As GPAI technology and market structures evolve, regulators should update compute thresholds and complement them with other metrics into regulatory review processes.
\end{abstract}

\newpage
\setcounter{footnote}{0}

\section*{Executive Summary}\addcontentsline{toc}{section}{Executive Summary}

\textbf{The development and deployment of advanced general-purpose
artificial intelligence (GPAI) models, also referred to as ``frontier AI
models'' or ``dual-use foundation models'', pose increasing risks of
large-scale societal harm (\cref{sec1:introduction}).} Currently, these
models develop ever higher capabilities through ever larger training
runs, fuelled by ever more computational resources (``training
compute''). But higher capabilities also mean higher risks to society,
because many capabilities are dual-use (e.g., automated hacking
capabilities) and because more capable models can be expected to be used
more widely and relied upon more heavily, increasing the stakes if they
fail or behave in undesired ways (e.g., producing biased outputs). As a
result, regulators are increasingly using training compute thresholds to
identify models of potential concern. 

\begin{figure}[ht]
    \centering
    \centerline{\includegraphics[width=1\linewidth]{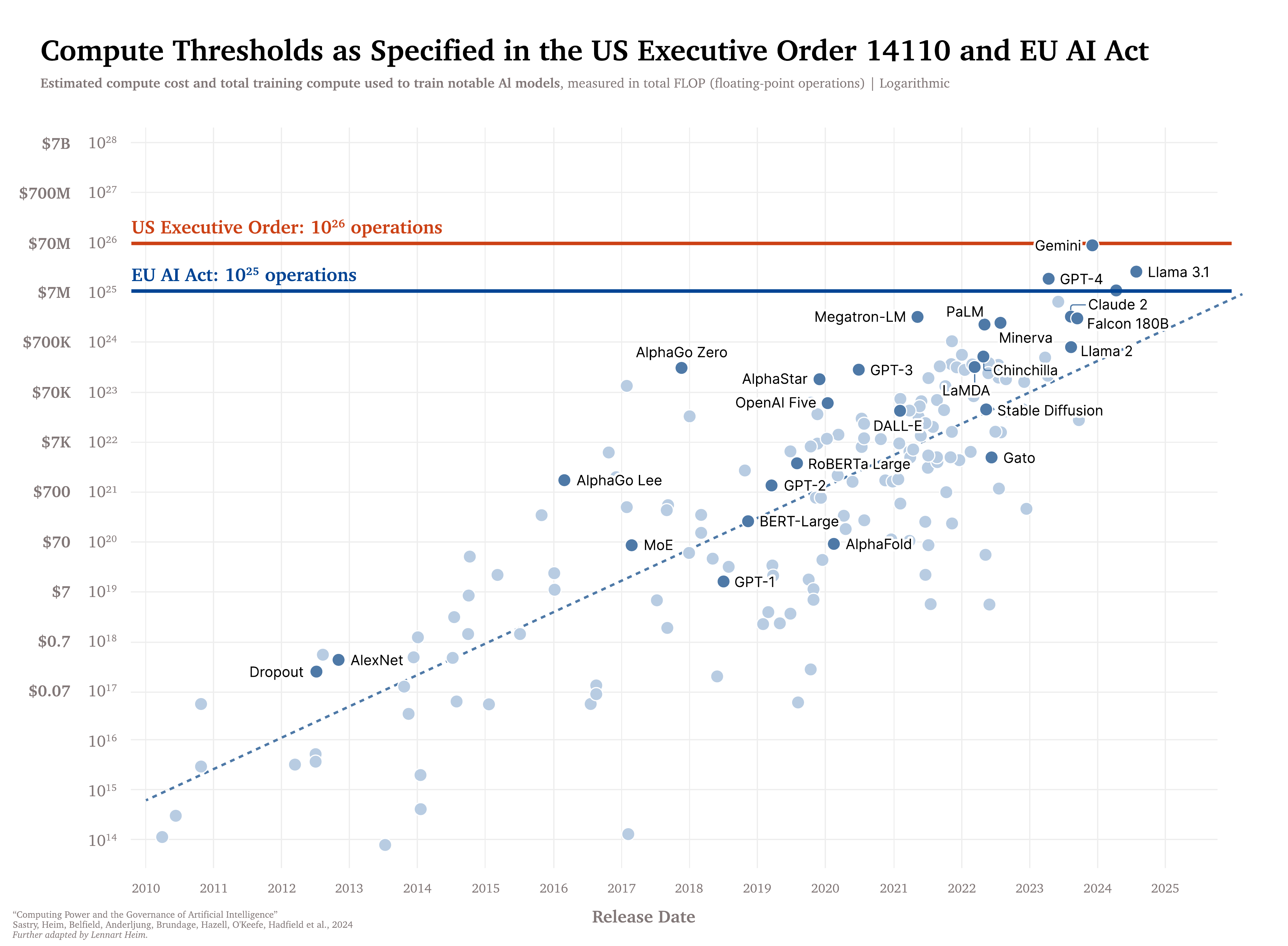}}
    \caption{Training compute has been increasing at a fast rate, doubling roughly every 6 months ($4\times$ per year). The US AI EO introduces reporting requirements for models trained with more than $10^{26}$ operations. The EU AI Act presumes a GPAI model poses systemic risk and imposes a variety of requirements for models trained with more than $10^{25}$ operations.}
    \label{fig:TKTK}
\end{figure}

\textbf{``Training compute'' refers to the total number of operations a
computer needs to perform to train an AI model (\cref{sec2:training-compute}).} In recent
years, the scale of AI training has grown significantly, with increases
in the amount of training data, the number of model parameters, and
corresponding increases in the amount of compute required for training (\Cref{fig:TKTK}). ``Compute'' in
this context refers to the total number of operations executed over the
training process. While post-training enhancements like fine-tuning can
significantly increase model capabilities, we recommend focusing on the
compute used for the large training run (``pre-training''), as this
aligns with empirical scaling laws and avoids impractical
re-measurements for successive fine-tuning instances.

\textbf{Training compute has several features useful for GPAI regulation
(\Cref{sec3:features-of-training-compute-useful-for-gpai-regulation}).
In particular, it is:}

\begin{itemize}
\vspace{-1em}
\item
  \textbf{Risk-tracking:} Training compute is indicative of a model's
  loss, capabilities, and risks. Empirical research has identified
  correlations, known as \emph{scaling laws}, between a
  model's training compute and its training loss, test
  loss, or validation loss. Improvements in loss tend to correlate with
  improvements in capabilities. As models become more capable, they may
  pose greater risks if they are misused or if they pursue misaligned
  objectives. The capabilities of a model also serve as a proxy for how
  widely it will be used and how heavily it will be relied upon and
  therefore the stakes if it fails or behaves in other undesired ways.

\item
  \textbf{Easily measurable:} Training compute is a quantifiable metric
  that is relatively simple and cheap to calculate.

\item
  \textbf{Difficulty of circumvention:} Training compute is relatively
  robust to circumvention attempts, as reducing the amount of compute
  used to train a model will generally decrease its capabilities and,
  consequently, lower its risks. This is because, for a given model
  architecture and training algorithm, the amount of compute used is
  directly related to the model's capabilities and potential risks. While algorithmic
  efficiency improvements gradually reduce the amount of training
  compute required for a certain level of performance over time, this
  represents an incremental progression of techniques rather than an
  active circumvention.

\item
  \textbf{Measurable before development and deployment:} Training
  compute can be calculated before the model is deployed, and estimated
  before the model is trained.

\item
  \textbf{Externally verifiable:} The possibility for external parties,
  such as compute providers, to verify compute usage, without disclosing
  proprietary details, enhances compliance.

\item
  \textbf{Cost-tracking:} Training compute is proportionate to the cost
  of computational resources for training, allowing the regulatory
  burden on smaller actors to be minimized while focusing on the most
  well-resourced ones.

\end{itemize}

\textbf{Notwithstanding the advantages discussed above, training compute
also has relevant limitations (\Cref{sec4:limitations-of-training-compute-relevant-for-gpai-regulation}).}
Fundamentally, it serves only as an imperfect proxy for risk. While high
training compute generally indicates increased model capabilities and
potential risks, some high-compute models may pose minimal concerns,
whereas certain low-compute models could present significant risks (with
the latter partially addressed through complementary non-GPAI
regulations). Moreover, improvements in algorithmic efficiency are
gradually reducing the amount of training compute required to achieve a
given level of capability, potentially weakening the relationship
between a specific compute threshold and the corresponding level of risk
over time. However, this shift in the compute-risk correlation might
occur gradually rather than abruptly as the field evolves.

\textbf{Consequently, we suggest that compute thresholds should be used
as an \emph{initial filter} to identify GPAI models that warrant
regulatory oversight and further scrutiny (\Cref{sec5.1:initial-filter}).} We argue
that compute thresholds can and should have two primary functions:
ensuring that regulators obtain sufficient visibility into risks of
large-scale societal harm from GPAI models and that companies invest
appropriate resources in understanding these risks. Precautionary
mitigation measures can be based on compute thresholds too, but it
should be allowable to later remove those mitigations if they turn out
to be unnecessary.

\begin{figure}[ht]
    \centering
   \centerline{\includegraphics[width=1\linewidth]{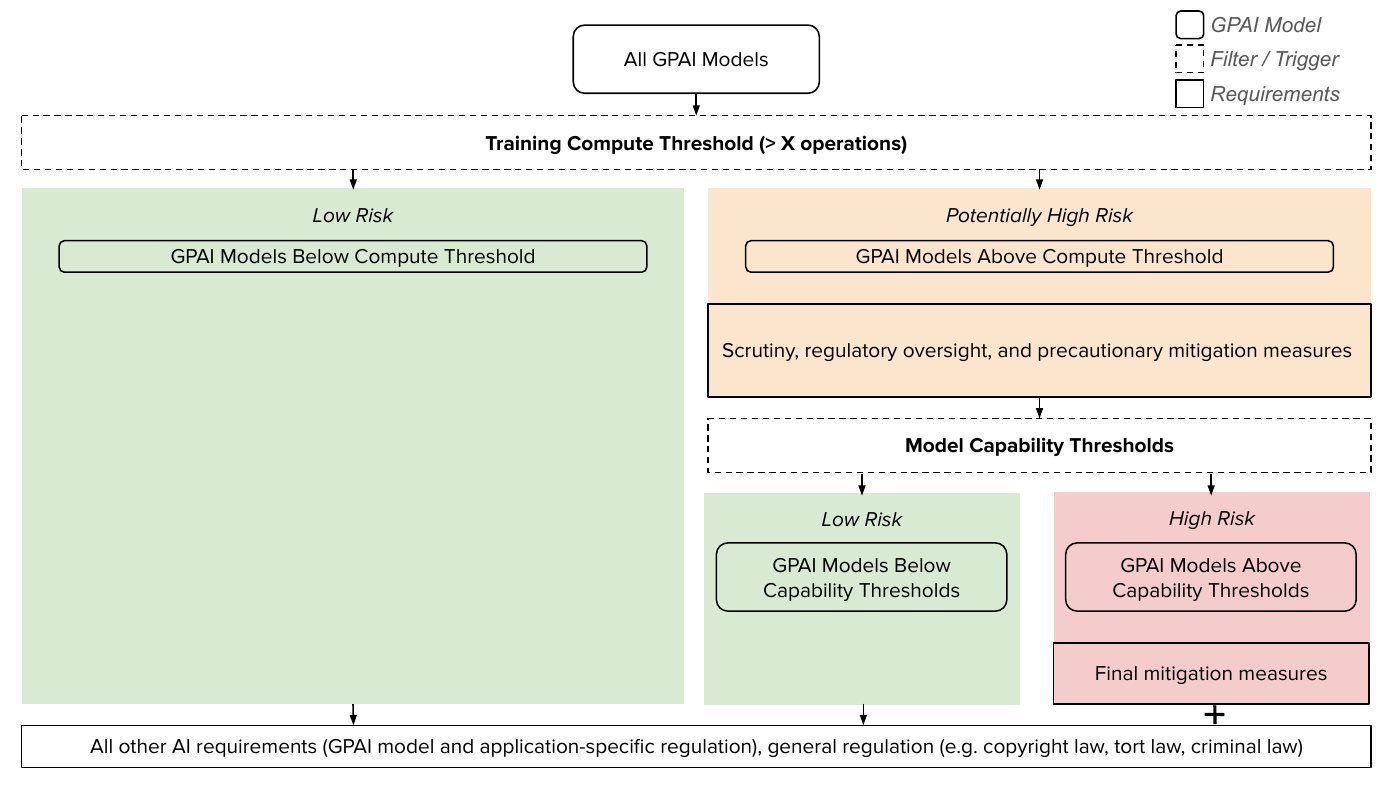}}
    \caption{Compute thresholds serve as an initial filter to identify GPAI models that warrant regulatory oversight and further scrutiny, and, for example, evaluation against capability thresholds to determine appropriate mitigation measures, complemented by other AI requirements.}
    \label{fig:TKTK2}
\end{figure}

\textbf{However, compute thresholds alone should generally \emph{not} determine
which mitigation measures are ultimately required}, given that compute is
only a crude proxy for model capabilities and an even cruder proxy for
risks of large-scale societal harm. Instead, to determine which
mitigation measures are required, compute thresholds can be complemented
with more precise but harder to evaluate thresholds based on other
metrics, such as capability thresholds based on model capability
evaluations. We also highlight that in a full regulatory framework for
AI, most requirements should not hinge on the amount of training compute (\Cref{fig:TKTK2}).

\textbf{Both the US AI EO and the EU AI Act mostly use compute
thresholds in line with our suggestions (\Cref{sec5.2:us-ai-eo} and \Cref{sec5.3:eu-ai-act}).} The US Executive Order 14110 mandates companies to notify the government about any ongoing or planned activities concerning the development of models that cross a compute threshold of $10^{26}$
operations, report on the measures taken to ensure the physical and
cybersecurity of model weights, and share the results of red-teaming
tests and mitigation measures taken based on those results. The \emph{EU
AI Act} requires providers of GPAI models that cross a compute threshold
of $10^{25}$ operations to notify the European Commission,
perform model evaluations, assess and mitigate systemic risks, report
serious incidents, and ensure cybersecurity of the model and its
physical infrastructure.

\textbf{Challenges remain in effectively leveraging compute thresholds
for GPAI regulation (\Cref{sec6:challenges-for-training-compute-thresholds-in-gpai-regulation}).}
A key question is the appropriate threshold level. There is high
uncertainty about the risk stemming from current and future GPAI models
trained on different amounts of compute. As a result, a low threshold
may be overinclusive, while a high threshold may be underinclusive.
Moreover, the direction in which a compute threshold should move over
time is not obvious, and it can make sense to complement compute
thresholds with other metrics.
For example,  increasing algorithmic efficiency allows more capable models to be trained with less compute. A reason to consider adjusting the compute threshold downwards over time.

However, there are also reasons to adjust the compute threshold upwards. If better
  understanding of GPAI models reveals that models that are in scope of
  a given compute threshold pose limited risks, the compute threshold
  should be raised to focus on potentially risky models. With improving
  algorithmic efficiency and computational price-performance, an
  increasing number of less well-resourced actors may fall within the
  scope of a given compute threshold, making the requirements
  increasingly burdensome and oversight increasingly costly. Other
  relevant factors include the threat landscape (i.e. the number,
  capacity, and willingness of malicious actors to use AI systems) and
  societal vulnerability or adaptation (i.e. the ability and capacity of
  society to deal with attacks, failures, and emergencies, e.g., through
  competent, well-resourced, and stable institutions).

Some other metrics, such as risk estimates, model
  capabilities, and effective compute, are better proxies for risk.
  However, these metrics are much harder to measure than training
  compute is. Complementing mere compute thresholds with other metrics
  becomes relevant to the extent that scaling laws cease to hold and
  training compute becomes a worse proxy for risk. Particularly relevant
  combinations may include training compute and model capability
  evaluations---to ensure catching the most capable models---and training
  compute and number of users---to ensure catching the most widely used
  models. However, any threshold that is supposed to serve as an
  initial filter to identify models of potential concern should be based
  on metrics that can be measured easily and early in the model
  lifecycle. This, at least currently, excludes risk estimates and model
  capability evaluations, and at least before deployment, the number of
  users.

\textbf{Overall, while not perfect, compute thresholds are currently a
key tool in GPAI regulation (\Cref{sec7:conclusion}).} In particular,
compute thresholds are currently the best tool available for identifying
potentially risky GPAI models and triggering regulatory oversight and
further scrutiny. They are based on a risk-correlated, quantifiable
metric that is difficult to circumvent and can be measured before model
development and deployment, enabling proactive governance efforts.
Compute thresholds can complement more targeted filters like model
capability evaluations that ultimately determine which mitigation
measures are required.

\clearpage

\tableofcontents

\clearpage

\section{Introduction}\label{sec1:introduction}

The development and deployment of advanced general-purpose artificial
intelligence (GPAI) models\footnote{There is no common distinction
  between what still counts as a ``model'' and what already constitutes
  a ``system.'' Regulators, such as the EU AI Office, need to provide
  clarification on this question. In this paper, we understand the term
  ``model'' to include model weights and source code, and the term
  ``system'' to encompass everything beyond (e.g., API, calls to other
  models, calls to other APIs) \citep[similar to][]{basdevant2024}.},
also referred to as ``frontier AI models'' or ``dual-use foundation
models'', pose increasing risks of large-scale societal harm
\citep{anderljung2023,bengio2024,hendrycks2023}. Currently, GPAI
models develop ever higher capabilities through ever larger training
runs, fuelled by ever more computational resources. The overall amount
of computational resources used to train a model is referred to as
\emph{training compute}. But higher capabilities also mean higher risks
to society, because many capabilities are dual-use (e.g., cyber
capabilities) and because more capable models can be expected to be used
more widely and relied upon more heavily, increasing the stakes if they
fail or behave in undesired ways (e.g., producing biased outputs). As a
result, regulators are increasingly imposing requirements on providers
of GPAI models whose amount of training compute passes a certain limit.
These limits are called \emph{compute thresholds}. This paper discusses
the key features of training compute and, accordingly, the functions
compute thresholds should have in GPAI regulation.

GPAI regulation worldwide increasingly relies on training compute
thresholds. Since October 2023, \emph{US Executive Order 14110 on the
Safe, Secure, and Trustworthy Development and Use of Artificial
Intelligence (US AI EO)} requires companies developing and deploying
GPAI models above a compute threshold of $10^{26}$ operations to notify the government, conduct red-teaming, and secure the
model weights (Section~4.2). From August 2024, the \emph{EU Regulation
2024/1689 laying down harmonised rules on artificial intelligence (EU AI
Act)} presumes that GPAI models that pass a compute threshold of
$10^{25}$ floating-point operations pose systemic risk
(Article~51(2)). Providers of such models are required to notify the
European Commission (Article~52(1)), conduct model evaluations, assess
and mitigate systemic risks, ensure adequate cybersecurity, and report
serious incidents (Article~55). Finally, an early draft of an
\emph{Artificial Intelligence Law of the People's Republic of China}
references training compute as one of the criteria used to identify
``critical AI'' which would require heightened safety and security
measures \citep{linghan2024}.\footnote{Throughout this paper, we use
  the terms ``(training) operations'', ``(training) compute'', and
  ``(training) compute threshold'' to emphasize that our focus is on the
  computational resources used specifically during the training phase of
  a AI model's lifecycle. There are other types of ``compute
  thresholds'' that can be used for AI regulation. For example, the US
  AI EO also includes a reporting requirement for computing clusters
  above a certain computing capacity per second. This is another type of
  ``compute threshold'' that does not directly refer to the training
  process but rather to the mere possession of computing infrastructure
  that can be used not only for training but also for deploying AI
  models.}

At the same time, the literature on compute thresholds is relatively
scarce. While many papers analyze trends in training compute \citep[e.g.,][]{lohn2022,pilz2023,sevilla2022,villalobos2024}, only a
few discuss using compute as a node for AI governance and suggest using
training compute thresholds \citep{anderljung2023,heim2024,egan2023,sastry2024}. \citet{hooker2024} questions training compute as a
proxy for risk and raises the issue that (training) compute thresholds
may quickly become obsolete. While the paper points at some relevant
challenges, as we argue in the following, it lacks nuance regarding the
features of training compute, the function that compute thresholds
\emph{can} and \emph{should} play in GPAI regulation, and an accurate
representation of regulations that already incorporate such thresholds.
The combination of increasing regulatory reliance on compute thresholds
and scarce academic literature suggests an urgent need for more
scholarly treatment of the topic.

This paper aims to answer the following two research questions: (1) What
features of training compute are relevant for GPAI regulation? (2) What
should be the function of (training) compute thresholds in GPAI
regulation? The remainder of the paper is structured as follows: \Cref{sec2:training-compute} clarifies the
concept of training compute,
\Cref{sec3:features-of-training-compute-useful-for-gpai-regulation}
discusses its useful features, and
\Cref{sec4:limitations-of-training-compute-relevant-for-gpai-regulation}
its limitations.
\Cref{sec5:functions-of-training-compute-thresholds-in-gpai-regulation}
suggests which functions (training) compute thresholds should have in
GPAI regulation, while
\Cref{sec6:challenges-for-training-compute-thresholds-in-gpai-regulation}
elaborates on challenges in that regard.
\Cref{sec7:conclusion} concludes with key
claims and further research questions.

\section{Training Compute}\label{sec2:training-compute}

Training an AI model is an iterative process where a model---a large
amount of numeric values (the so-called ``parameters'') arranged in a
certain way (the so-called ``architecture'')---is exposed to a large
amount of data, allowing the model to learn from the data by adapting
the parameters. This learning can be supervised, where the model is
provided with labeled examples, or unsupervised, where the model learns
from unlabeled data \citep{goodfellow2016}.

``Training compute'' or simply ``compute'' refers to the amount of
computational resources or, more precisely, the total number of
operations required to train an AI model. The training process runs on a
computer that performs a vast number of mathematical operations. We are
agnostic here as to whether these are integer operations, floating-point
operations (FLOP), or other operations. Presently, most AI training
predominantly uses floating-point operations, but this could change in
the future. We therefore broadly refer to all types of operations as
``(training) operations (OP).'' The total number of operations used to train
the model is a quantity we refer to as ``(training)
compute.''\footnote{When referring to the quantity of computational
  resources used during training, it is preferable to use the term
  ``FLOP'' rather than ``FLOPs.'' For a longer elaboration, see
  \citet{heim2023}.}

The term ``compute'' in this context refers to the number of operations
executed in total over the whole training process, not to be confused
with the computer's processing performance---the number of operations
the computer is able to execute per second (FLOP/s or OP/s). For
example, running an NVIDIA A100 processor (which has a processing
performance of 312 TeraFLOP/s for FP16 tensor) for a week results in a
total of $1.9\times 10^{20}$ FLOP being executed. Multiplied with
the number of NVIDIA A100 processors used, this results in the total
number of operations used to train the AI model.\footnote{Note that this
  is a simplified explanation. For more details, see \citet{pistilloforthcoming}.}

\begin{figure}[ht!]
    \centering
    \centerline{\includegraphics[width=1.2\linewidth]{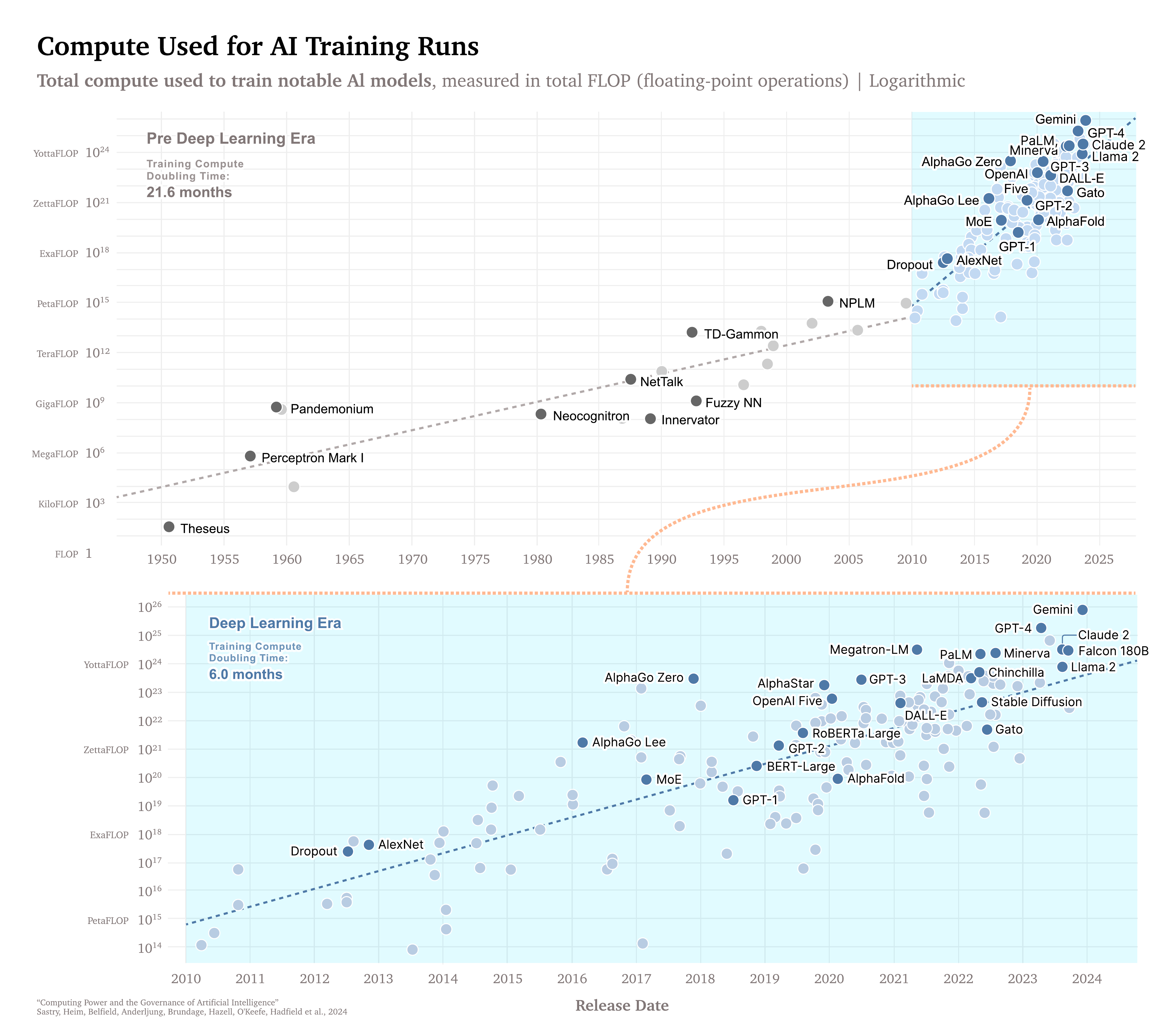}}
    \caption{Amount of compute used to train AI models over time. In the pre-deep learning era, training compute followed Moore's Law, doubling approximately every two years. Since the emergence of the Deep Learning Era around 2010, training compute has been increasing at a much faster rate, doubling roughly every 6 months (increasing by about $4\times$ per year). This rapid growth is largely driven by increased investments in computational resources for training larger models, which have demonstrated improved capabilities (figure from \citealt{sastry2024}; up-to-date as of end of 2023; underlying data and updates can be found at \citealt{2024}).}
    \label{fig:fig1}
\end{figure}

In recent years, the scale of AI training has grown significantly, with
increases in the amount of training data, the number of model
parameters, and, therefore, the amount of compute required for training
(growing by about $4\times$ every year since 2010,
\Cref{fig:fig1}).

There is currently no standardized method for measuring a model's
training compute. This introduces ambiguity that may be considered
problematic in a regulatory context. We suggest measuring training
compute by following the guidance of the Frontier Model Forum
\citep{frontiermodelforum2024}. More precisely, we suggest using
Method 1 described in \citep{sevilla2022a}. However, given that the amount of training compute used for the final large training run
(``pre-training'') currently increases by about four times per year
\citep{epochai2024,sevilla2022}, the exact details of these methods
are not critical.

\begin{figure}[ht!]
    \centering
    \centerline{\includegraphics[width=1\linewidth]{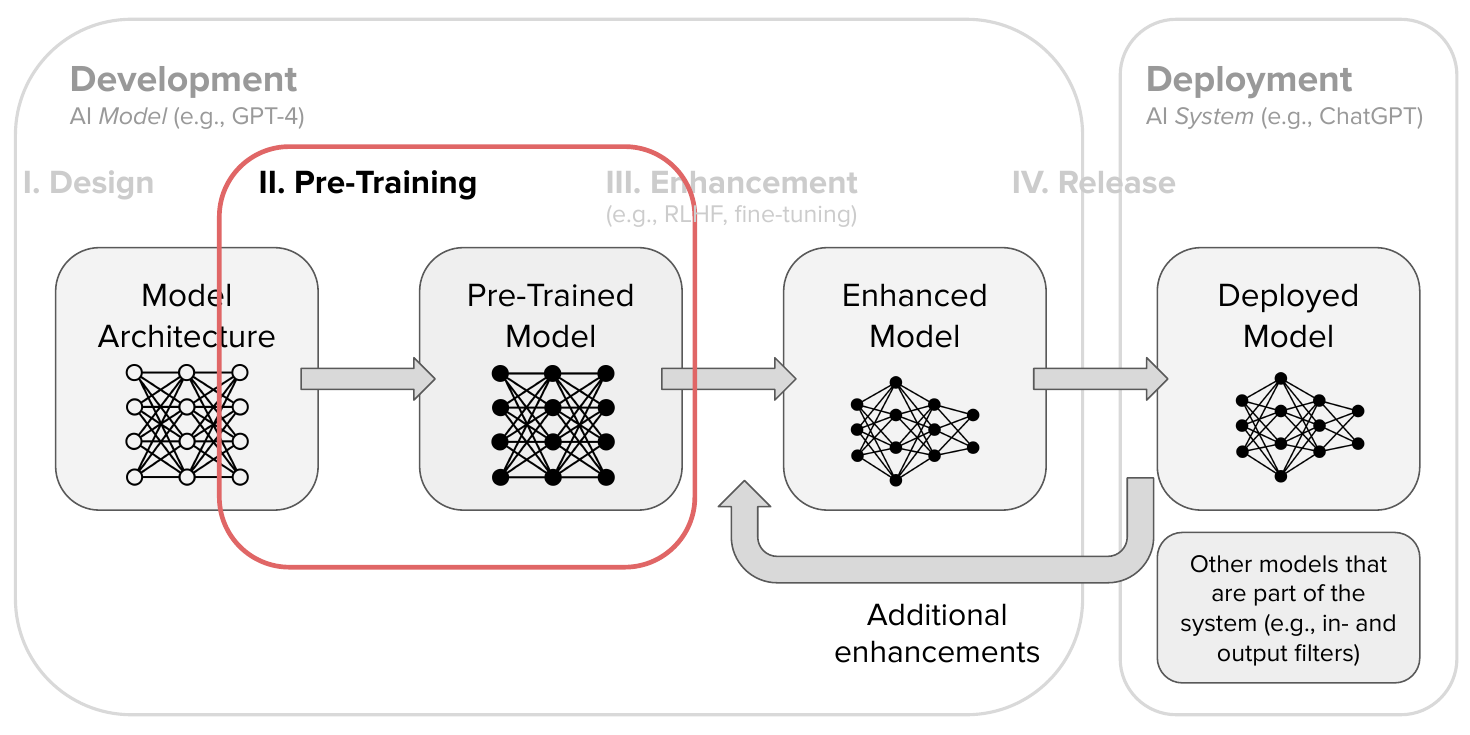}}
    \caption{We recommend only measuring pre-training compute and
not including compute used in further enhancement processes (figure
adapted from \citet{pistilloforthcoming}).}
    \label{fig:fig2}
\end{figure}

By contrast, an important question is whether to focus on the compute
used for pre-training alone or whether to also include the compute used
for ``fine-tuning'', ``reinforcement learning from human feedback''
(RLHF), and other post-training enhancements (\Cref{fig:fig2}). Post-training
enhancements can significantly improve model capabilities, up to an
equivalent of a $5\times$ to $20\times$ increase in training compute
\citep{davidson2023,villalobos2023}. Presumably to account for this,
Recital 111 of the EU AI Act suggests including compute used for
post-training enhancements: ``\emph{The cumulative amount of computation
used for training includes the computation used across the activities
and methods that are intended to enhance the capabilities of the model
prior to deployment, such as pre-training, synthetic data generation and
fine-tuning.}''

We believe that including the compute used for post-training
enhancements lacks an empirical basis, is unnecessary, and is
impractical. First, the empirical basis for existing scaling laws
describes model performance as a function of pre-training compute, not
of compute used for fine-tuning or other post-training enhancements (for
more on scaling laws, see \Cref{sec3:features-of-training-compute-useful-for-gpai-regulation}).
This means there is no empirical basis for imposing requirements based
on the cumulative amount of compute. Second, fine-tuning compute is
small compared to pre-training compute.\footnote{Usually around 1\%, or, in the largest case we are aware of, up to about 14\% of the pre-training compute. A small robotics model (Swift) is an exception, using fine-tuning compute equivalent to about 200\% of pre-training compute \citep{2024}.} Consequently, including fine-tuning compute
would make only a small difference to the overall training compute
measurement. Lastly, measuring fine-tuning compute is impractical
because fine-tuning is repeated many times for a given pre-trained model
and often performed by downstream developers. Re-measuring training
compute (and potentially re-reporting and re-conducting model
evaluations) for every fine-tuned version would be extremely burdensome
for individual regulatees and lead to a much larger number of
regulatees.

Instead, we recommend only measuring
pre-training compute. While post-training enhancements can significantly
increase a model's capabilities, this should not be considered directly
when designing a compute threshold, but only indirectly as discussed in \Cref{sec6.1:where-to-set-compute-thresholds}. Moreover, post-training enhancements can and should be taken into account when measuring a model's capabilities, as it is the enhanced
model that will be deployed in the end. Indeed, comprehensive model
evaluations should involve applying fine-tuning and other post-training
enhancements to elicit the pre-trained model's full capabilities as much as possible. However, it is not necessary or practical to re-conduct
model capability evaluations on every enhanced model version. Minor variations may
not necessitate re-conducting model capability evaluations, provided the
pre-trained model has already been evaluated. We suggest re-conducting
model capability evaluations on enhanced model versions periodically and if they
constitute a new model family.

\section{Features of Training Compute Useful for GPAI
Regulation}\label{sec3:features-of-training-compute-useful-for-gpai-regulation}

Training compute has several features that make it a valuable metric for
GPAI regulation. Namely, training compute is indicative of a model's
capabilities and therefore its risks, easily measurable, difficult to
circumvent, measurable before development and deployment, externally
verifiable, and indicative of the developer's resources and therefore
its capacity to handle regulatory burdens. In the following, we discuss
each of these features in more detail.

\begin{enumerate}[wide]
\item
  \textbf{Risk-tracking: Training compute is indicative of a model's
  loss, capabilities, and risks.}

  \begin{itemize}
  \item
    For GPAI models, the amount of compute used to train a model
    correlates with the model's loss, capabilities, and risks it may
    present. Empirical research has identified correlations, known as
    \emph{scaling laws}, between a model's training compute and its
    training loss, test loss, or validation loss \citep{hernandez2021,
    hoffmann2022,kaplan2020,sutton2019,
    epoch2023scalinglawsliteraturereview}. Further, improvements in
    loss tend to correlate with improvements in capabilities on
    downstream tasks \citep{brown2020,ganguli2022,rae2022}.
    Finally, as models become more capable, they may pose greater risks
    if they are misused or if they pursue misaligned objectives. The
    capabilities of a model also serve as a proxy for how widely it will
    be used and how heavily it will be relied upon and therefore the
    stakes if it fails or behaves in other undesired ways.\footnote{At
      the same time, the risks from failure and other undesired
      behaviors decrease with higher capabilities in terms of model
      accuracy, robustness, and not being susceptible to prompt
      injection attacks and jailbreaks. Which one of these two trends
      outweighs the other is currently impossible to say---empirical
      data is lacking to make statements about the past, let alone
      predict the future.} For example, this may include increases in
    the number and severity of people affected by biased outputs
    \citep{bommasani2022}.
  \item
    This feature of training compute---its correlation with a model's
    loss, capabilities, and risks---is the most important but also
    perhaps the most controversial \citep[e.g.,][]{hooker2024}. While
    this feature is already disputed today, we highlight that to the
    extent that scaling laws cease to hold in the future---for example,
    because a training paradigm other than deep learning emerges---
    training compute will become a less useful metric for GPAI
    regulation. We discuss this feature in some more depth in \Cref{sec4:limitations-of-training-compute-relevant-for-gpai-regulation}.
  \end{itemize}
\item
  \textbf{Easily measurable: Training compute is a quantifiable metric
  that is relatively simple and cheap to calculate.}

  \begin{itemize}
  \item
    Training compute is a metric that is easy to measure, as it can be
    directly calculated from model specifications or inferred from data
    about the use of hardware with minimal effort. Training compute also
    is unidimensional and durable, unlike other metrics, many of which
    are multidimensional (e.g., data quality and type) or may quickly
    become outdated (e.g., model capability benchmarks).
  \end{itemize}
\item
  \textbf{Difficult to circumvent: Training compute is difficult to
  reduce without also decreasing a model's capabilities and risks.}

  \begin{itemize}
  \item
    Training compute is relatively robust to circumvention attempts, as
    reducing the amount of compute used to train a model will generally
    decrease its capabilities and, consequently, lower its risks. This
    is because, for a given model architecture and training algorithm,
    the amount of compute used is directly related to the model's
    capabilities and potential risks. Therefore, a GPAI developer cannot
    simply decide to use less compute while maintaining the same level
    of capabilities. In contrast, a GPAI developer might be able to
    adjust other metrics, such as the score on a specific benchmark, to
    avoid regulation without substantially impacting the model's
    capabilities.
  \item
    However, improvements in algorithmic efficiency can reduce the
    amount of compute required for a given level of capabilities. Over
    time, this may allow less well-resourced actors to develop models
    that achieve a given level of capabilities \citep{pilz2023a}.
    Indeed, improvements in algorithmic efficiency pose a challenge to
    using compute thresholds. However, while future algorithmic
    efficiency improvements may require less compute for the same level
    of capabilities, this represents natural progress rather than active
    circumvention attempts. Purposefully decreasing training compute for
    fixed algorithms would still come at the cost of lower capabilities
    and risks. We address the question of when to update compute
    thresholds to account for algorithmic efficiency improvements in \Cref{sec6.3:why-when-and-how-to-update-compute-thresholds}.
  \end{itemize}
\item
  \textbf{Measurable before development and deployment: Training compute
  can be calculated before the model is deployed and even before it is
  trained.}

  \begin{itemize}
  \item
    Training compute can be known ahead of deployment and estimated
    ahead of development. This is important because regulators may want
    to impose requirements for how a model is developed and deployed.
    Training compute can be calculated before model deployment because
    training will be completed at that time. Before model development,
    training compute can be estimated using the architectural details
    and the amount of training data, as outlined in Method 1 of
    \citet{sevilla2022a}. AI companies carefully plan their training
    runs, as training state-of-the-art models often requires tens of
    thousands of GPUs and costs on the order of millions of dollars.
    Given the significant computational resources and financial
    investment involved, companies have a strong incentive to accurately
    estimate training compute beforehand to ensure efficient resource
    allocation and budget planning. By estimating training compute
    before training begins, developers can implement compute-indexed
    precautions during the training process. For example, they can
    ensure that strong cybersecurity measures are in place for
    compute-intensive training runs, reducing the risk of model theft or
    unauthorized access.
  \end{itemize}
\item
  \textbf{Externally verifiable: The possibility for external parties to
  verify compute usage, without disclosing proprietary details, enhances
  compliance.}

  \begin{itemize}
  \item
    Ideally, measurements of training compute are verifiable by diverse
    external parties through protocols that maintain the confidentiality
    of proprietary information. This could also enable verifiable
    commitments across companies and even states \citep{brundage2020}.
    Compute providers can aid in the verification of requirements based
    on training compute. This is particularly desirable, as compute
    providers can monitor and verify compute usage without infringing on
    the confidentiality of developers, in contrast to model capabilities
    and other metrics that may require access to sensitive model
    details. This is because compute usage can be monitored and verified
    without the need to access specific details of the model
    architecture, training data, or other proprietary information
    \citep{heim2024}.
  \end{itemize}
\item
  \textbf{Cost-tracking: Training compute is proportionately higher for
  models that cost more to develop, minimizing the regulatory burden on
  smaller actors while focusing on the most well-resourced ones.}

  \begin{itemize}
  \item
    The amount of compute used to train a model directly corresponds to
    the amount of financial resources required to do so (\Cref{fig:fig3}). In other
    words, the cost of training a model scales with the amount of
    training compute used. For example, training a model with
    $10^{26}$ operations (\$70M) will cost approximately 10
    times more than training one with $10^{25}$ operations
    (\$7M).
  \item
    The large amounts of compute required to train state-of-the-art
    models are typically only available to well-resourced organizations.
    By setting compute thresholds at appropriate levels, regulators can
    focus on the most advanced and potentially risky models without
    imposing undue burdens on smaller actors in the AI ecosystem, such
    as startups, small businesses, or academic researchers.
  \item
    However, the cost of a given amount of compute decreases over time.
    Computational price-performance (FLOP per \$) has been observed to
    double every 2.1 years for machine learning GPUs and 2.5 years for
    general GPUs \citep{hobbhahn2023}. As the cost of compute falls,
    more actors may be able to develop models that cross the compute
    threshold \citep{pilz2023a}. This expansion in the regulatory
    scope would impose regulatory burdens on smaller actors and pose a
    challenge in terms of scaling oversight as the number of covered
    entities grows. In \Cref{sec6.3:why-when-and-how-to-update-compute-thresholds},
    we discuss approaches to account for a regulatory scope that is
    potentially expanding due to improving computational
    price-performance.
  \end{itemize}
\end{enumerate}

\begin{figure}
    \centering
    \centerline{\includegraphics[width=1.2\linewidth]{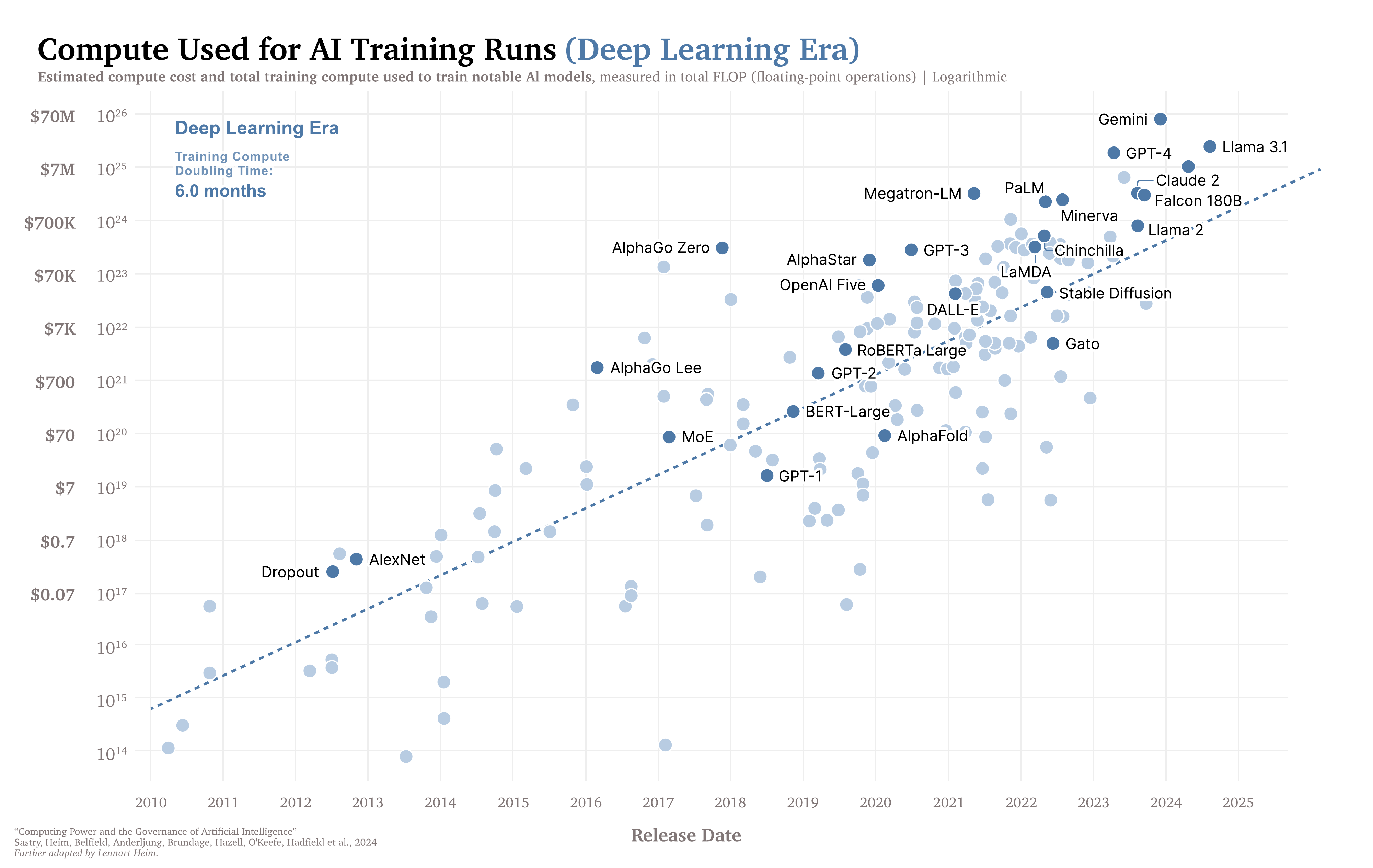}}
    \caption{Cost and compute used for training AI models. The amount of compute used to train a model directly corresponds to the amount of financial resources required to do so.\protect\footnotemark This rapid growth is largely driven by increased investments in computational resources for training larger models, which have demonstrated improved capabilities (figure adapted from \citealt{sastry2024}; underlying data and updates can be found at \citealt{2024}).}
    \label{fig:fig3}
\end{figure}
\footnotetext{While the exact cost calculations vary across different sources due to uncertainty about potential discounts and the achieved utilization or efficiency of leveraging the provided computing power \citep[see][]{hobbhahn2023}, the rough estimates tend to fall within a similar range. For the purposes of this analysis, we assume a cost of \$1.8 per hour for renting an NVIDIA H100, which is a rather optimistic assumption. Also, note that this is only the cost of acquiring the computational resources for the final large training run. It does not include staff cost and compute usage beyond pre-training.}

\section{Limitations of Training Compute Relevant for GPAI
Regulation}\label{sec4:limitations-of-training-compute-relevant-for-gpai-regulation}

In this section, we discuss the main limitations of training compute
relevant for frontier AI regulation. We argue that training compute is
only a very crude proxy for risk and that, in the future, training
compute may become a worse proxy for risk or even cease to be such a
proxy altogether.

Scaling laws describe relationships between training compute and a
model's loss (\Cref{sec3:features-of-training-compute-useful-for-gpai-regulation}).
The subsequent relationship between loss and capabilities is not always
``smooth.'' For example, inverse scaling has been observed for some
tasks \citep{koralus2023,mckenzie2024,perez2022}, though for some
of these tasks, the relationship has been re-established with further
scaling \citep{openai2023,wei2022}. Moreover, with gradual
improvements in loss, there can be jumps in capabilities, or at least
apparent jumps depending on how granular the capabilities are measured
\citep{wei2023,wei2022a}.\footnote{These apparent discontinuities
  arise because we often care about discontinuous measures, such as
  getting the exact right answer in mathematics or avoiding crashes in
  autonomous vehicles, rather than just approximating the correct
  solution. While continuous surrogate measures could potentially be
  used to forecast when such discontinuous jumps in capability might
  occur, identifying the appropriate continuous measure is itself a
  complex problem. For a more comprehensive discussion of ``emergent
  capabilities'', see
  \citet{anderljung2023}, \citet{pistilloforthcoming}, and \citet{woodside2024}.}

The relationship between increases in capabilities and risks is also not
``smooth.'' The higher a model's capabilities, the more risky the model
will arguably be, because it can be misused for more dangerous purposes
and it is likely to be employed in more and higher-stakes settings.
However, risk is highly contextual. Factors other than model
capabilities have a major impact on risk. Important factors include the
offense-defense balance of AI systems, i.e. whether AI systems help
more with beneficial or harmful uses\footnote{The offense-defense
  balance of AI systems could significantly impact how increased model
  capabilities translate to risks. If defensive AI applications can
  reduce AI risks, it may counterbalance risks from offensively used
  models.} \citep{perspectives2024,buterin2023,lohn2022a}, the
threat landscape, i.e. the number, capacity, and willingness of
malicious actors to use AI systems \citep{koessler2024}, and
societal vulnerability or adaptation, i.e. the ability and capacity of
society to deal with attacks, failures, and emergencies, e.g., through
competent, well-resourced, and stable institutions
\citep{anderljung2023,bernardi2024,kapoor2024}.

Essentially, training compute correlates with loss, which correlates
with capabilities, which in turn correlate with risk. Thus, training
compute is a proxy for risk; however, as none of these three
correlations are perfect, training compute can only be considered a very
crude proxy for risk \citep{hooker2024}.\footnote{For a discussion of the relationship between training compute,
  loss, capabilities, and risks, see \citet{anderljung2023},
  \citet{sastry2024}, and \citet{pistilloforthcoming}.}

Fundamentally, the notion of training compute as a proxy for risk hinges
on the validity of scaling laws. While scaling laws are empirical
observations derived from past data, they do not guarantee that these
relationships will persist indefinitely. There is a prediction, referred
to as the ``scaling hypothesis'', that these relationships will continue
to hold true in the future \citep{branwen2020,sutton2019}. However,
innovation in AI might gradually lead to a paradigm shift away from deep
learning, or the pre-training phase might become less crucial for
determining the final capabilities of AI models. As a result, training
compute may become a less precise proxy for risk over time or could
potentially cease to serve as an effective proxy for risk altogether.
Nonetheless, we do not anticipate an abrupt disruption in the
relationship between training compute and risk. Any potential shift is
likely to occur gradually as the field evolves over an extended period.
In \Cref{sec6.3:why-when-and-how-to-update-compute-thresholds},
we discuss approaches to account for changes in the relationship between
training compute and risk due to improving algorithmic efficiency.

\section{Functions of Training Compute Thresholds in GPAI
Regulation}\label{sec5:functions-of-training-compute-thresholds-in-gpai-regulation}

Building on the characteristics of training compute outlined in the
previous sections, in this section we argue that compute thresholds can
and should be used as an initial filter to identify models that warrant
regulatory oversight, further scrutiny, and precautionary safety
measures. However, given that compute is only a very crude proxy for
capabilities and risk, compute thresholds should not be the final
determinant for what safety measures to require. After compute
thresholds, decision criteria based on more precise proxies for the risk
a model poses, such as capability thresholds, should be applied (\Cref{sec5.1:initial-filter}). Both the US AI EO (\Cref{sec5.2:us-ai-eo})
and the EU AI Act (\Cref{sec5.3:eu-ai-act}) take this approach.

\subsection{Initial Filter}\label{sec5.1:initial-filter}

When imposing requirements on GPAI models, policymakers need to consider
five interdependent questions:

\begin{enumerate}
\item
  Which \emph{risks} should be countered? (E.g., large-scale societal
  harm.)
\item
  Which \emph{models} could pose these risks? (E.g., advanced GPAI
  models.)
\item
  Which \emph{metrics} correspond to features of these models? (E.g.,
  training compute.)
\item
  Which \emph{thresholds} are the values of those metrics at which
  models start to pose these risks? (E.g., $10^{26}$ operations of training compute.)
\item
  Which \emph{requirements} should be imposed on models above those
  thresholds to counter these risks? (E.g., reporting, model evaluation,
  and risk assessment.)
\end{enumerate}

Easily measurable metrics, such as training compute, are generally
preferable to ensure legal certainty about which GPAI models are in
scope of requirements. At the same time, metrics that are better proxies
for the risk a GPAI model poses, such as model capabilities, are
generally preferable to ensure requirements are targeted at risky GPAI
models. There will often be a trade-off between ease of measurability
and correlation with risk (more in \Cref{sec6.4:alternative-metrics}). Taken
together, metrics that are cruder proxies for the risk a GPAI model
poses should be combined with less costly and less definitive
requirements, while metrics that are better proxies for the risk a GPAI
model poses may be linked to more costly and more definitive
requirements. What is more, all else equal, the stricter the
requirements, the higher and therefore more exclusive the threshold
should be.

Building on these considerations, the function of compute thresholds
should be to serve as an \emph{initial filter} to identify models of
potential concern. Training compute is an easily measurable and
externally verifiable proxy for risk that allows focusing on the most
risky models and the most well-resourced actors (\Cref{sec3:features-of-training-compute-useful-for-gpai-regulation}).
Therefore, compute thresholds provide an easy way to identify those GPAI
models that warrant heightened attention (both from companies and
regulators), while filtering out the much larger fraction of GPAI models
that are highly unlikely to pose risks of large-scale societal harm. As
a result, using compute thresholds as an initial filter can help to
reduce both compliance burdens for regulatees and enforcement costs for
regulators, while focusing attention on those GPAI models that are most
likely to pose risks of large-scale societal harm.

However, the function of compute thresholds should \emph{only} be to
serve as an initial filter to identify models that warrant regulatory
oversight and further scrutiny. Already, training compute is a very
crude proxy for risk, and this relationship may become worse in the
future (\Cref{sec4:limitations-of-training-compute-relevant-for-gpai-regulation}).
Therefore, compute thresholds should be complemented by thresholds based
on metrics that are more precise but that may be harder to evaluate,
such as model capability evaluations. In particular, compute thresholds
should not be used to ultimately determine which mitigation measures
need to be taken (beyond a few key precautionary mitigation measures).
For models that cross a compute threshold, more rigorous analyses of the
risk they pose should be conducted through model evaluations and risk
assessments. Based on the results of these analyses, specific mitigation
measures can be required (and precautionary mitigation measures that
have been taken but turn out unnecessary may be discontinued).
Currently, a sensible approach would be to first apply compute
thresholds (to identify GPAI models that warrant oversight and scrutiny)
and to subsequently apply capability thresholds (to ultimately determine
which mitigation measures need to be implemented) (\Cref{sec4:limitations-of-training-compute-relevant-for-gpai-regulation}).

\begin{figure}[ht!]
    \centering
    \centerline{\includegraphics[width=1.2\linewidth]{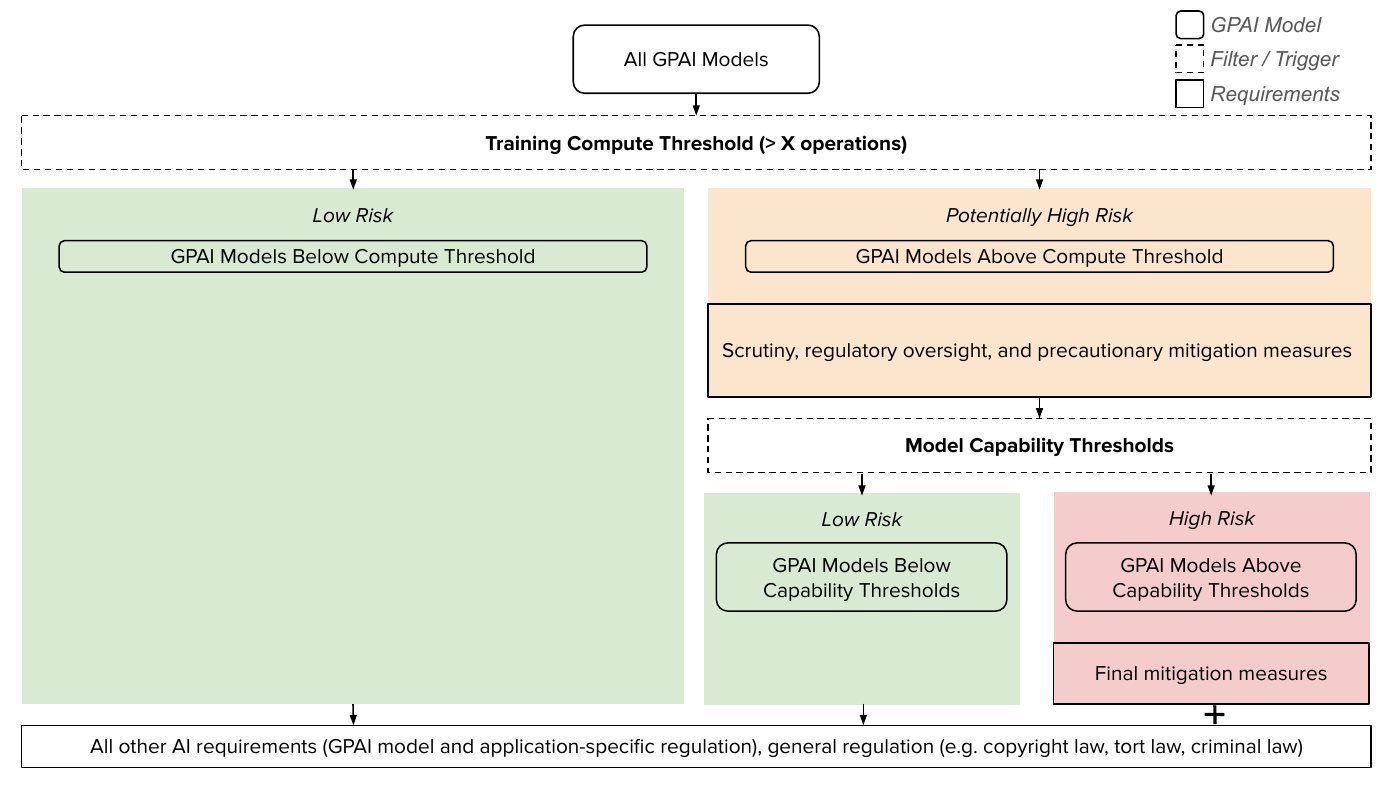}}
    \caption{A framework for the role of compute thresholds in GPAI regulation. Compute thresholds serve as an initial filter to identify GPAI models that warrant regulatory oversight and further scrutiny, and, for example, evaluation against capability thresholds to determine appropriate mitigation measures, complemented by other AI requirements.}
    \label{fig:fig4}
\end{figure}

We also highlight that compute thresholds should not be used for all
requirements aimed to counter risks from AI. First, many risks from AI
stem from the context and way in which AI models are applied, instead of
or in addition to the intrinsic properties of those AI models,
necessitating requirements at the application layer. In particular,
bias, discrimination, and fairness risks, for example, caused by AI
models being applied in education, hiring, and public service provision,
should be countered through requirements at the application layer in
addition to the model layer, as done in the EU AI Act Chapter II
(Prohibited AI Practices), Chapter III (High-Risk AI Systems), and
Chapter IV (Transparency Obligations). For requirements at the
application layer, compute thresholds will usually not make sense but
other criteria should be decisive, like the type of application. Second,
some risks from GPAI models arise regardless of their amount of training
compute, such as risks of copyright or privacy infringements. These
risks should be countered through requirements imposed on all GPAI
models, as done in the EU AI Act Article~53. Third, ex post regulation
like tort law and criminal law should of course not hinge on compute
thresholds. Overall, in a full regulatory framework for AI, most
requirements should not hinge on the amount of training compute (\Cref{fig:fig4}).

Thresholds based on easily measurable metrics that serve as initial
filters for regulatory oversight and further scrutiny are in place
across various sectors. For example, in the US, the Environmental
Protection Agency's (EPA) ecological risk assessments start by
determining whether some concentration of contaminants has been crossed,
only after which a more thorough risk assessment takes place
\citep{u.s.environmentalprotectionagency1996}. In the EU, the Digital
Services Act (DSA) identifies ``very large online platforms'' and ``very
large online search engines'' that need to conduct systemic risk
assessments based on whether they have a ``number of average monthly
active recipients of the service in the Union equal to or higher than 45
million'' (Article~33(1) DSA). Similarly, the General Data Protection
Regulation (GDPR) mandates data protection impact assessments for
entities processing data ``on a large scale'' (Article~35(3) GDPR)---
the concrete metric and threshold in this case is to be specified by
member states, who have chosen different thresholds, most of which are
based on the number of people whose data is concerned
\citep{breitbarth2018}.

While we generally have reservations about using compute thresholds to
ultimately determine which mitigation measures to require, this can be
different if the threshold is set relatively high such that only a few
well-resourced actors are affected. For example, we have suggested
elsewhere to use compute thresholds to identify large compute providers
on which to impose Know Your Customer (KYC) and other requirements
\citep{egan2023}. We believe this approach can be justified, as it
only targets a handful of companies \citep[see][]{heim2024,egan2023}.

\subsection{US AI Executive Order}\label{sec5.2:us-ai-eo}

The US AI EO covers a range of issues from AI. Section~4 leverages
training compute thresholds as a criterion for classifying AI models
that warrant regulatory oversight and further scrutiny due to potential
safety and security concerns. Specifically, it targets:

\begin{quote}
\emph{``any model that was trained using a quantity of computing power
greater than $10^{26}$ integer or floating-point
operations, or using primarily biological sequence data and using a
quantity of computing power greater than $10^{23}$ integer
or floating-point operations''} (Section~4.2(b)(i))
\end{quote}

\begin{figure}[ht!]
    \centering
    \centerline{\includegraphics[width=1.2\linewidth]{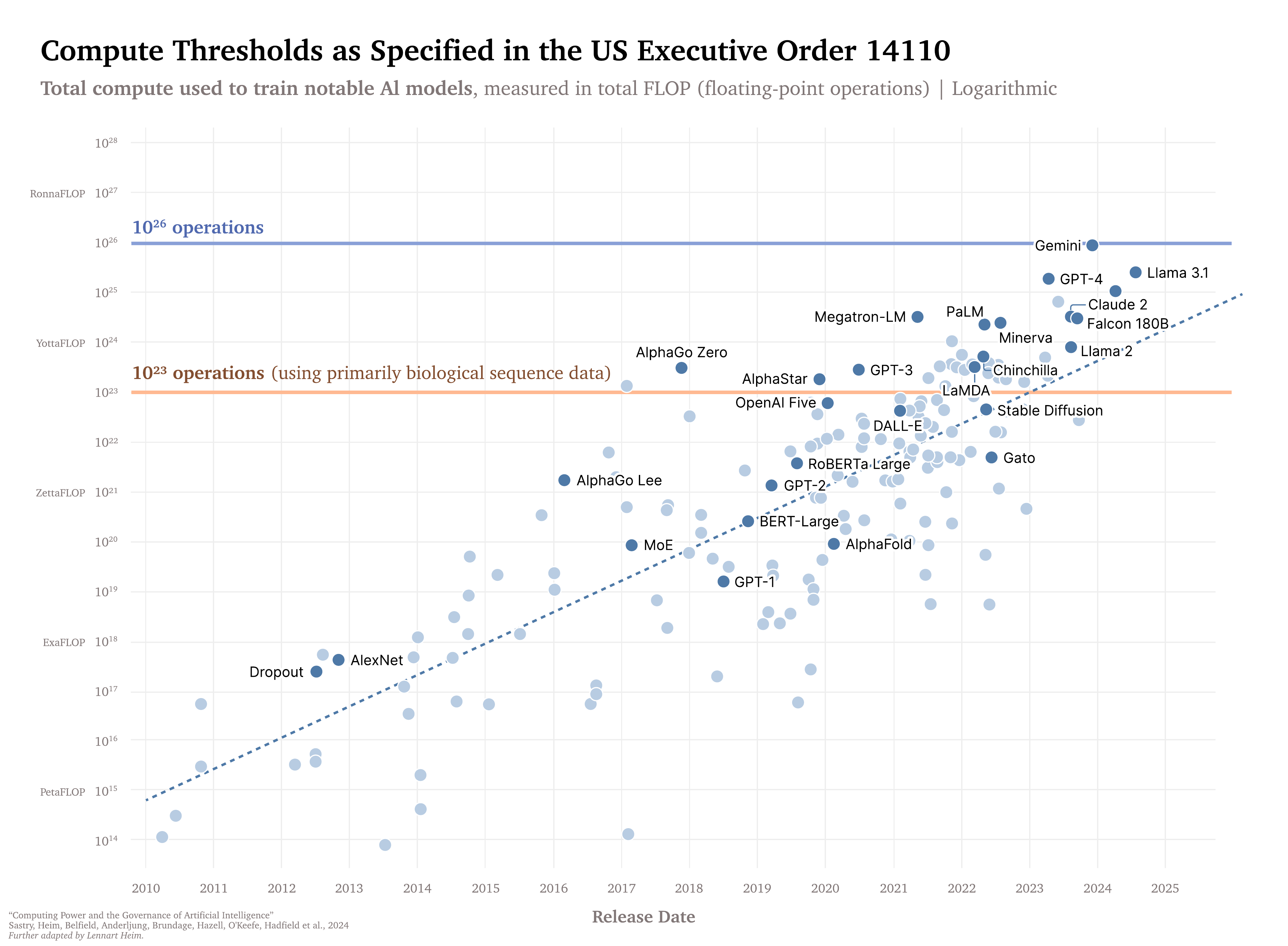}}
    \caption{The US AI EO introduces reporting requirements for models trained with more than $10^{26}$ operations and $10^{23}$ operations if trained using primarily biological sequence data (figure adapted \citealt{sastry2024}; underlying data and updates can be found at \citealt{2024}).}
    \label{fig:fig5}
\end{figure}

Note that these thresholds are designed to capture future models. As of
July 2024, no model has been officially reported to meet the general
compute threshold of $10^{26}$ operations (\Cref{fig:fig5}). However, one model
has been estimated to meet the lower bio-compute threshold
\citep{maug2024}.

The US AI EO mandates companies to notify the government about ongoing
or planned activities concerning the development of models that cross
the compute thresholds (Section~4.2(a)(i)(A)). It also requires these
companies to report on the measures taken to ensure the physical and
cybersecurity of model weights (Section~4.2(a)(i)(B)) and share the
results of red-teaming tests and mitigation measures taken based on
those results (Section~4.2(a)(i)(C)).

The US AI EO uses compute thresholds in line with what we have argued
can and should be their role (\Cref{sec5.1:initial-filter}). Its
requirements focus on reporting to increase regulatory oversight, which
is one of the primary functions compute thresholds can fulfill. While
the US AI EO does not mandate companies to implement security measures
or conduct red-teaming tests, the corresponding reporting requirements
strongly push companies to take such measures in order to not appear
irresponsible to the regulator. In our view, the pressure to conduct
red-teaming tests also makes sense, as increased scrutiny is another
primary function of compute thresholds. The pressure to implement
security measures can also be based on compute thresholds. However, we
believe that companies should be given the option to relax their
security measures again if they establish that their model does not
exceed a specific level of capabilities or can otherwise be proven to be
sufficiently safe. We discuss domain-specific compute thresholds in \Cref{sec6.2:domain-specific-compute-thresholds}.

\subsection{EU AI Act}\label{sec5.3:eu-ai-act}

The EU AI Act is the most comprehensive regulatory framework on AI
worldwide, prohibiting certain AI practices (Chapter II), imposing
requirements on certain high-risk AI systems (Chapter III), requiring
transparency about the use of AI in services and to produce content
(Chapter IV), and imposing requirements on GPAI models (Chapter V).
Regarding the latter, the EU AI Act leverages a compute threshold to
distinguish between GPAI models with and without systemic risk.
Concretely, it draws the line in the following way:\footnote{When
  establishing compute thresholds, it is advisable for the EU AI Act to
  use the term ``operations'' rather than specifying a particular type
  of operation (such as floating-point operations or FLOP). Using the
  broader term ``operations'' ensures that the threshold remains
  agnostic to the specific type of computational operations performed
  during training, making it more future-proof in light of possible
  technological changes. This approach also maintains consistency with
  the terminology used in the US AI EO, ensuring interoperability and
  enabling cooperation between EU and US regulators.}

\begin{quote}
\emph{``A general-purpose AI model shall be presumed to have high impact
capabilities pursuant to paragraph 1, point (a) {[}and thus be
classified as posing systemic risk{]}, when the cumulative amount of
computation used for its training measured in floating point operations
is greater than $10^{25}$.''} (Article~51(2))
\end{quote}

\begin{figure}[ht!]
    \centering
    \centerline{\includegraphics[width=1.2\linewidth]{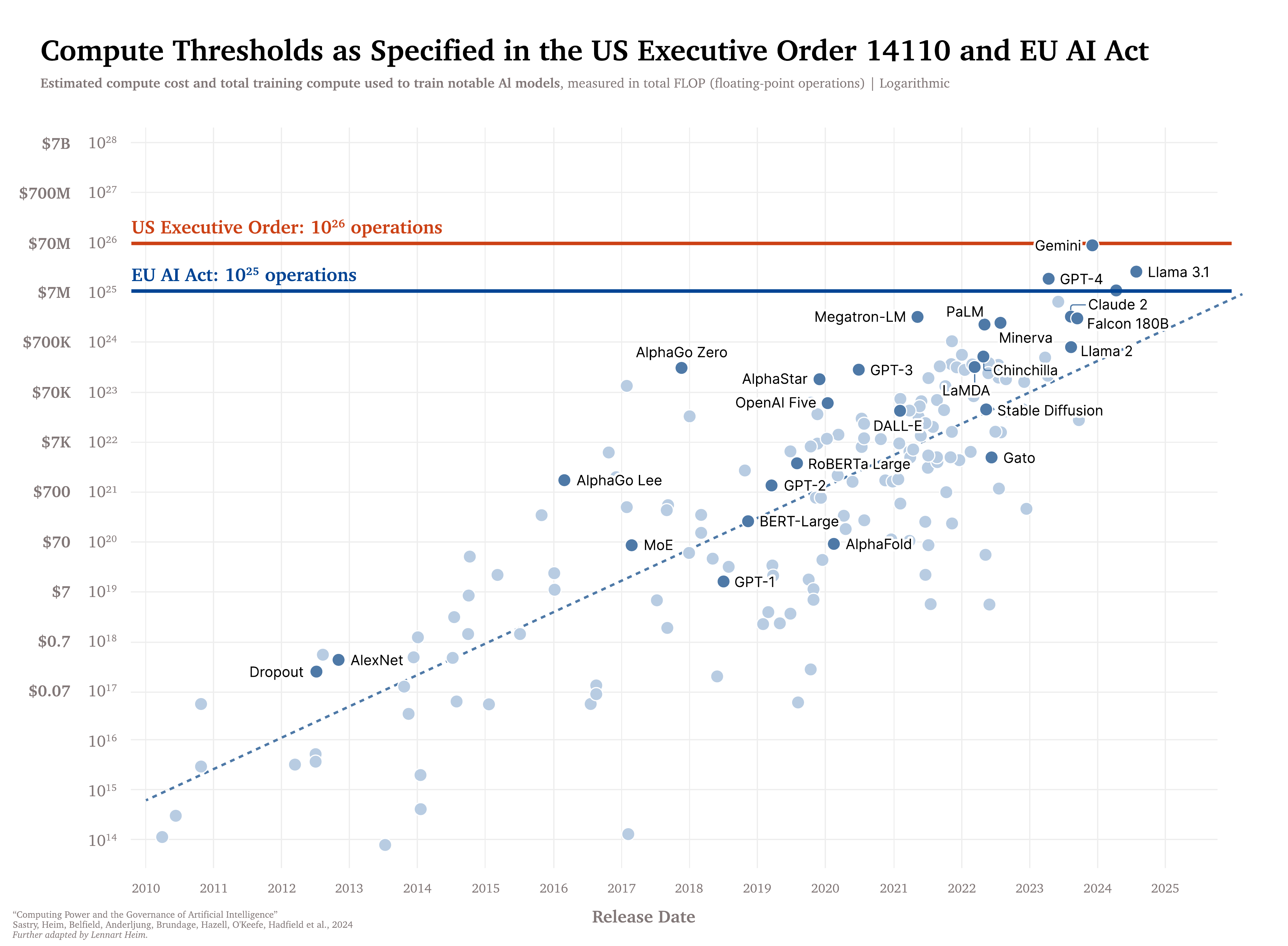}}
    \caption{While the US AI EO introduces reporting requirements for models trained with more than $10^{26}$ operations, the EU AI Act presumes a GPAI model poses systemic risk and imposes a variety of requirements for models trained with more than $10^{25}$ operations (figure adapted from \citealt{sastry2024}; underlying data and updates can be found at \citealt{2024}).}
    \label{fig:fig6}
\end{figure}

In contrast to the general compute threshold in the US AI EO, this
compute threshold has immediate relevance. As of July 2024, one model
available in the EU has officially been reported to have crossed the
$10^{25}$ training compute threshold (Inflection-2), some
additional models available have been estimated to have done so (e.g., Gemini
Ultra, GPT-4, Inflection-2.5) \citep{2024}, and there may be a handful more of such models (\Cref{fig:fig6}).

\begin{figure}[ht!]
    \centering
    \centerline{\includegraphics[width=1.2\linewidth]{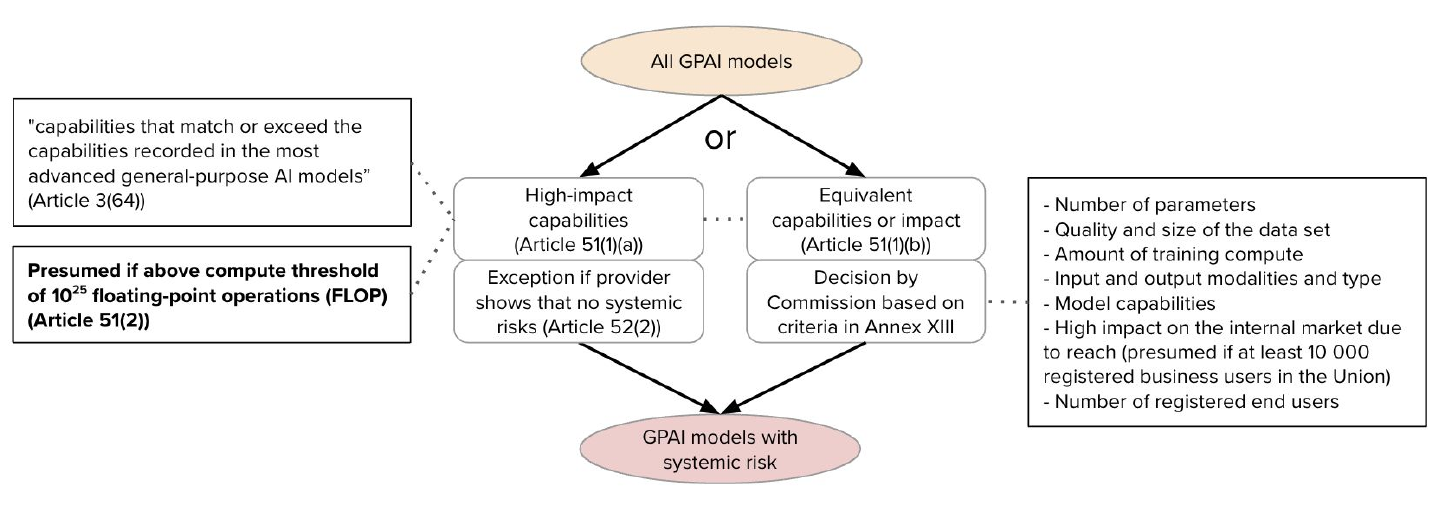}}
    \caption{The EU AI Act outlines two main pathways for classifying general-purpose AI (GPAI) models as posing systemic risk. The first path deems models as high-risk if they possess ``high-impact capabilities'' that match or exceed the most advanced GPAI models, which is presumed if the training compute exceeds $10^{25}$ floating-point operations (FLOP). Alternatively, the European Commission can classify models based on criteria outlined in Annex XIII. Models deemed high-risk through either pathway are subject to stringent regulations to mitigate systemic risks.}
    \label{fig:fig7}
\end{figure}

In the EU AI Act, the compute threshold is only one of several ways to
identify GPAI models with systemic risk, but it currently is the most
concrete one. In more detail, Article~51(1) says that a GPAI model shall
be classified as posing systemic risk if it has either ``high impact
capabilities'' (Article~51(1)(a)) or ``capabilities or an impact
equivalent to those set out in point (a)'' (Article~51(1)(b)). Regarding
the first alternative, high-impact capabilities are defined as
``capabilities that match or exceed the capabilities recorded in the
most advanced general-purpose AI models'' (Article~3(64)). A model is
\emph{presumed} to have high-impact capabilities, and thus pose systemic
risk, if the compute threshold of $10^{25}$ floating-point
operations is crossed (Article~51(2)). Regarding the second alternative,
it is not defined and remains somewhat unclear what ``capabilities or an
impact equivalent to those set out in point (a)'' means. Instead, the
European Commission decides whether these conditions are fulfilled
``having regard to the criteria set out in Annex XIII''
(Article~51(1)(b)). These criteria contain many of the metrics discussed
in \Cref{sec6.4:alternative-metrics},
including the number of parameters, the quality and quantity of the
training data, and the number of users (Annex XIII). Overall, the
compute threshold currently seems to be the most concretely outlined way
in which a GPAI model can be classified as posing systemic risk, and may
therefore be the most relevant in the near term (\Cref{fig:fig7}).

Article~52(1) requires providers to notify the European Commission if a
GPAI model crosses or will cross the threshold laid out in
Article~51(1)(a), in particular by passing the compute thresholds in
Article~51(2). This is to be done ``without delay and in any event
within two weeks.'' Furthermore, Article~55 requires providers of GPAI
models with systemic risk to perform model evaluations, assess and
mitigate systemic risks, report serious incidents, and ensure
cybersecurity of the model and its physical infrastructure.\footnote{These
  requirements apply in addition to the requirements for all GPAI
  models, which include documentation, transparency, and copyright
  policies (Article~53).} These requirements are relatively vague but
will be concretized by ``codes of practice''
(Article~56).\footnote{It is very attractive, though not mandatory, for
  providers to comply with the codes of practice, because this provides
  a ``presumption of conformity.'' If a provider complies with the codes
  of practice, the provider is presumed to also comply with the EU AI
  Act. But providers have the option to demonstrate compliance in other
  ways, too (Article~55(2)).} \footnote{Ultimately,
  technical standards are supposed to provide the presumption of
  conformity (Article~40(1)). However, the European Commission has not
  yet issued the standardization request for GPAI models, and once this
  happens it will still take several months or years for the technical
  standard to be developed. Therefore, in the meantime, the codes of
  practice take on the equivalent role.}

The EU AI Act uses compute thresholds mostly in line with what we have
argued can and should be their role (\Cref{sec5.1:initial-filter}). It focuses on
requirements of regulatory oversight (notification and serious incidents
reporting) and further scrutiny (model evaluations and systemic risk
assessments). The requirements to mitigate systemic risks and ensure
adequate cybersecurity may appear to go beyond what we think compute
thresholds should be used for---we have argued that they should not be
used to determine which mitigation measures need to be implemented. But
this is not necessarily how these requirements need to be interpreted.
The codes of practice can and, in our view, should differentiate further
which mitigation measures must be implemented under which circumstances
(for example, based on the results of model evaluations), such that this
is not solely determined by the compute threshold.

\begin{figure}[ht!]
    \centering
    \centerline{\includegraphics[width=1.2\linewidth]{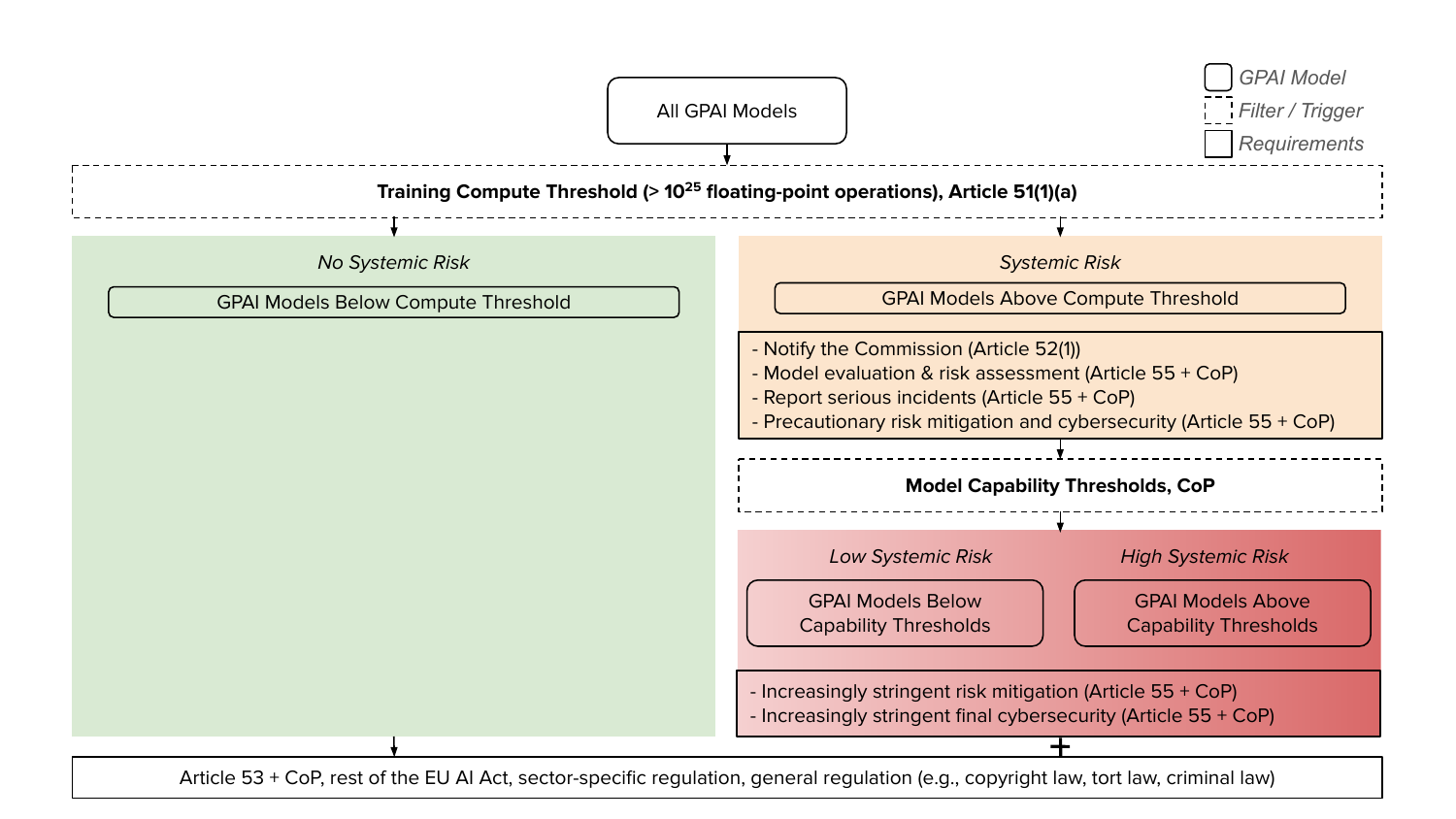}}
    \caption{How the EU AI Act can be understood to implement the framework of \Cref{fig:fig4}. GPAI models exceeding the $10^{25}$ FLOP training compute threshold (Article 51(1)(a)) are classified as posing systemic risk, triggering requirements such as notifying the Commission, undergoing model evaluations and risk assessments, reporting incidents, and implementing precautionary risk mitigation and cybersecurity measures. Subsequent model capability thresholds, outlined in Codes of Practice (CoP), further categorize these high-risk models into low or high systemic risk levels, informing increasingly stringent mitigation and cybersecurity requirements.}
    \label{fig:fig8}
\end{figure}

Another important mechanism in the EU AI Act is the possibility for
providers to demonstrate that their model does not pose systemic risk,
despite fulfilling the conditions of Article~51(1)(a)---such as passing
the compute threshold (Article~52(2)). This exception especially makes
sense if the conditions of Article~51(1)(a) are presumed to be fulfilled
because the compute threshold is crossed, given that training compute is
only a very crude proxy for risk (\Cref{sec4:limitations-of-training-compute-relevant-for-gpai-regulation}).
The provision is in line with how we have argued compute thresholds
should be used
(\Cref{sec5.1:initial-filter}). If it turns
out that a model does not pose systemic risk despite passing the compute
threshold---for example, based on model evaluations and risk
assessments---providers will be allowed to refrain from further model
evaluations, risk assessments, etc., and drop any precautionary security
measures they have already taken. Overall, the EU AI Act can be
understood to embed its compute threshold in a framework similar to what
we have suggested in \Cref{sec5.1:initial-filter} (\Cref{fig:fig8}).

\section{Challenges for Training Compute Thresholds in GPAI
Regulation}\label{sec6:challenges-for-training-compute-thresholds-in-gpai-regulation}

The key question when using compute thresholds is at what level to place
the threshold. In \Cref{sec6.1:where-to-set-compute-thresholds},
we discuss where a compute threshold should be set at a given point in
time. In \Cref{sec6.2:domain-specific-compute-thresholds},
we briefly touch on whether domain-specific compute thresholds make
sense. In \Cref{sec6.3:why-when-and-how-to-update-compute-thresholds},
we examine why, when, and how a compute threshold should be updated over
time. In \Cref{sec6.4:alternative-metrics},
we discuss metrics other than training compute that may be useful to
distinguish between more and less risky models, with a focus on but not
limited to risk estimates, model capability evaluations, and effective
compute.

\subsection{Where to Set Compute
Thresholds}\label{sec6.1:where-to-set-compute-thresholds}

In this section, we discuss at what level a compute threshold should be
set. Already, different regulators have chosen different values, with
the US AI EO using $10^{26}$ operations and the EU AI Act
using $10^{25}$ floating-point operations, one order of
magnitude lower. This difference is relevant, as the EU threshold likely
captures several existing models, while the US threshold probably does
not capture any existing model (\Cref{sec5:functions-of-training-compute-thresholds-in-gpai-regulation}).

In general, at any given point in time, a compute threshold can capture
three tiers of models:

\begin{enumerate}
\def\labelenumi{\arabic{enumi}.}
\item
  \textbf{Models above the frontier} (currently
  $10^{26}$--$10^{27}$ operations). At any
  given point in time, such models are the most important ones to
  include, as models above the frontier may have unprecedented and
  hard-to-predict dangerous capabilities \citep[see][]{anderljung2023}.
  Better understanding of and government visibility into
  state-of-the-art models is key to identifying potential issues as
  early as possible, as well as for getting a sense for what may be
  coming down the pike. The advantage of this approach is that it
  focuses on the few models that are the most likely to be risky. On the
  other hand, it incurs some chance of being underinclusive, missing
  risky models.
\item
  \textbf{Models at the frontier} (currently
  $10^{25}$--$10^{26}$ operations).
  Currently, including such models may be deemed warranted, as models at
  the current frontier may already be considered to show warning signs
  of some dangerous capabilities \citep[see][]{anthropic2024,
  openai2023,phuong2024}. While some companies developing such
  models have reported that their models do not possess dangerous
  capabilities, governments may want to verify those claims and ensure
  that all companies developing such models conduct model evaluations
  and risk assessments. This seems to be the approach of the EU AI Act.
\item
  \textbf{Models below the frontier} (currently
  $10^{24}$--$10^{25}$ operations).
  Currently, including such models would be a very cautious approach.
  There still is some uncertainty about the full extent of the
  capabilities of models below the current frontier \citep[see][]{anderljung2023}. The advantage of this approach is that it
  creates a large ``margin of safety.'' On the other hand, it has the
  biggest chance of being overinclusive, putting regulatory burdens on
  companies developing models that later turn out to be not risky. This
  is burdensome for regulators, if they want to properly verify company
  reports, and for companies themselves. However, in the future, if
  society does not keep pace in adapting to increasingly high model
  capabilities, maintaining a threshold below the frontier could be
  warranted (\Cref{sec6.3:why-when-and-how-to-update-compute-thresholds}).
\end{enumerate}

Currently, the US AI EO, at least for now, has taken the first approach,
only targeting models above the current frontier. Whether the US will
uphold this approach by updating its compute thresholds with the moving
frontier remains to be seen (\Cref{sec6.3:why-when-and-how-to-update-compute-thresholds}).
The EU AI Act has explicitly taken the second approach, targeting models
at and above the current frontier: A GPAI model poses systemic risk if
it has ``high impact capabilities'' (Article~51(1)(a)), which in turn
are defined as ``capabilities that match or exceed the capabilities
recorded in the most advanced general-purpose AI models''
(Article~3(64)). This arguably means the provision aims to refer to a
moving set of models---those at and above the frontier \emph{at any
given point in time}. If the provision was referring to the models at
and above the frontier \emph{at a specific point in time}, such as the
time of writing or entering into force of the EU AI Act, this point in
time would have been specified for legal certainty. Based on this
interpretation, the compute threshold in Article~51(2) would need to be
updated with the moving frontier (see also \Cref{sec6.3:why-when-and-how-to-update-compute-thresholds}).

\subsubsection*{Post-Training Enhancements}\label{post-training-enhancements}

An important phenomenon that should affect where to set compute
thresholds is the possibility of post-training enhancements (\Cref{sec2:training-compute}). Compute thresholds should be based on the capabilities of enhanced
\emph{systems} like ChatGPT, rather than the capabilities of pre-trained
models, because those are the capabilities the models will ultimately
obtain. For example, if a regulator currently sets a threshold at
$10^{25}$ operations, this should be because they have
concluded that models like GPT-4 or Gemini Ultra 1.0 and their
superseeding application, such as ChatGPT or Gemini, should be subject
to regulation. This assessment should not be based on the capabilities
of the pre-trained model but rather on those achieved through
post-training enhancements. By contrast, if a regulator currently sets a
threshold at $10^{26}$ operations, this should be because
they believe that no current model possesses concerning capabilities,
even after accounting for post-training enhancements.

Regulators could also add a safety buffer of an order of magnitude or so
to account for potential future improvements in post-training
enhancements. For pre-trained models trained on the same amount of
compute, better post-training enhancements could be developed over time.
Future improvements in post-training enhancements could lead to models
trained on a given amount of pre-training compute being enhanced to
display higher and more dangerous capabilities than current models
trained on the same amount of pre-training compute but enhanced with
current techniques only. To account for further unknown post-training
enhancements, regulators could add a safety buffer in ``equivalent
pre-training compute'' \citep{davidson2023}.

However, we expect that for the foreseeable future there is likely a
limit to the increase in model capabilities that can be gained from
post-training enhancements. The capabilities of the model determined by
pre-training will likely remain critical. While post-training
enhancements may continue to improve incrementally, they are unlikely to
enable models to ``leapfrog'' or ``skip multiple generations'' of compute
scaling and achieve capabilities comparable to those of models
pre-trained with orders of magnitude more compute
\citep{davidson2023}. Furthermore, certain post-training enhancements
work better with larger models \citep{li2024}, suggesting their
effectiveness may depend on leveraging the capabilities of models with
significant pre-training compute. In other words, while post-training
enhancements can potentially enhance the absolute capabilities of
models, they are less likely to significantly alter the relative
capabilities compared to models pre-trained with orders of magnitude
more compute, especially for more advanced models.

\subsection{Domain-Specific Compute
Thresholds}\label{sec6.2:domain-specific-compute-thresholds}

The focus of this paper up to this point has been on compute thresholds
for GPAI models. This section briefly discusses domain-specific compute
thresholds for narrower, more specialized models. Already, the US AI EO
includes a compute threshold for models specialized in the domain of
biology. The considerations for whether it makes sense to use
domain-specific compute thresholds in general, and in the domain of
biology in particular, are the same as the considerations for whether it
makes sense to use compute thresholds for GPAI models.

Generally, the key consideration is whether scaling laws apply in the
domain in question. Regulating models based on the amount of training
compute only makes sense if there is a strong relationship between the
amount of compute used to train a model and the level of risk the model
poses. Another important consideration is how many actors will be in
scope of a given compute threshold. If too many actors are in scope,
i.e., the amount of computational power to surpass the threshold is
available to a broad range of actors, a domain-specific compute
threshold may not be practicable. Both of these considerations may
differ from domain to domain.

We are uncertain whether compute thresholds in the domain of biology
make sense and can only provide initial thoughts: In the domain of
biology, a distinction can be made between AI-enabled biological tools
(BTs) and large language models (LLMs) with a focus on biology
\citep{sandbrink2023,smith2024}. AI-enabled BTs are AI models that
are highly specialized on a particular biological task, such as
predicting protein structures, like AlphaFold 2 \citep{jumper2021}.
LLMs with a focus on biology are LLMs trained on a lot of biological
sequence data, and can be used for a variety of applications, such as
ESM3 \citep{hayes2024}. It is uncertain if compute thresholds are
appropriate for AI-enabled BTs, as the literature on scaling laws is not
well established for those. In contrast to GPAI models, other factors,
such as the quality of data, are much more important for their
capabilities. Moreover, AI-enabled BTs generally require about $100\times$ less
compute to train than GPAI models, such that a large number of actors
would be in scope of requirements based on compute thresholds
\citep{smith2024}. For LLMs with a focus on biology, compute
thresholds could make sense, but the literature on scaling laws is
similarly scarce (e.g., \citet{hesslow2022,serrano2024}).

\subsection{Why, When, and How to Update Compute
Thresholds}\label{sec6.3:why-when-and-how-to-update-compute-thresholds}

Compute thresholds will very likely need to be updated over time.
Generally, it is important to maintain flexibility in regulations for a
poorly understood and rapidly evolving technology like GPAI. First and
foremost, compute thresholds should be adjusted with better knowledge
about what constitutes risky models. What is more, depending on
improvements in algorithmic efficiency and computational
price-performance, as well as changes in the offense-defense balance of
AI systems, the threat landscape, and societal vulnerability or
adaptation, compute thresholds will need to be adjusted downwards or
upwards over time.

The US AI EO states that its thresholds for models in scope should
regularly be updated:

\begin{quote}
\emph{``The Secretary of Commerce, in consultation with the Secretary of
State, the Secretary of Defense, the Secretary of Energy, and the
Director of National Intelligence, shall define, and thereafter update
as needed on a regular basis, the set of technical conditions for models
and computing clusters that would be subject to the reporting
requirements of subsection~4.2(a) of this section.''} (Section~4.2(b))
\end{quote}

The EU AI Act also states that its thresholds for GPAI models with
systemic risk should be updated based on technological developments:

\begin{quote}
\emph{``The Commission shall adopt delegated acts in accordance with
Article~97 to amend the thresholds listed in paragraphs 1 and 2 of this
Article, as well as to supplement benchmarks and indicators in light of
evolving technological developments, such as algorithmic improvements or
increased hardware efficiency, when necessary, for these thresholds to
reflect the state of the art.''} (Article~51(3))
\end{quote}

Given that GPAI is a poorly understood and rapidly evolving technology,
any GPAI regulation should have the ability to be updated. On a high
level, there are two approaches to updating the level of compute
thresholds over time: specifying a fixed value but tasking regulators
with updating that value over time, or specifying a dynamic value that
automatically moves according to certain rules over time. Both the US
and the EU have specified fixed values but tasked regulators with
updating those values over time. By contrast, for example, a compute
threshold could specify a dynamic value that automatically moves over
time by being tied to the highest-compute model at any given point in
time, or, alternatively, by being set at an order of magnitude below the
highest-compute model at any given point in time. But even a dynamic
compute threshold should allow for updates by regulators to account for
a better understanding of what constitutes risky models as well as for
technological and other relevant developments.

The frequency and the direction of updates to training compute
thresholds depend on several developments. First and foremost, over
time, our understanding of what constitutes risky models may improve.
Currently, uncertainty about model capabilities and societal risks is
high. Regulators impose requirements on GPAI models not because they are
sure those models pose a high level of risk, but because they might. As
such, if the current approach is motivated by precaution and has the
goal to detect when GPAI models reach some level of risk, the compute
threshold can be expected to move up over time as our understanding of
model capabilities and societal risks improves.

Moreover, compute thresholds may need to be adjusted downwards over time
to account for algorithmic efficiency improvements. In the context of
model training, algorithmic efficiency refers to the amount of compute
required to run a given algorithm used to train a model. Over time,
ongoing research yields improved model architectures and training
algorithms, reducing the amount of compute needed to train models of a
given level of capabilities \citep{pilz2023a}. Adjusting the compute
threshold downwards at the pace of algorithmic efficiency improvements
would maintain its correspondence to a given level of capabilities.
However, algorithmic efficiency improvements can also mean that more
actors will be able to train models with a given amount of compute and
thus obtain a given level of capabilities. This can lead to high
regulatory burdens on a large number of less well-resourced actors like
startups and academics, and, correspondingly, high oversight costs for
regulators \citep{pistilloforthcoming}. As a result, algorithmic
efficiency improvements can also provide an argument to adjust the
compute threshold upwards over time. Taken together, algorithmic
efficiency improvements may pose a regulatory dilemma \citep{christophwinter}.

At the same time, algorithmic efficiency improvements merely define the
upper limit for how quickly the compute threshold may need to be
adjusted downwards to apply to models posing a given level of risk.
Despite improvements in algorithmic efficiency, developments such as the
offense-defense balance of AI systems, the threat landscape, and
societal vulnerability or adaptation suggest the threshold may need
slower downwards adjustments, remain unchanged, or even be adjusted
upwards to catch models of a given level of risk. GPAI models may turn
out to enable defensive uses less, as much as, or even more than
offensive uses, for example, in the cyber domain
\citep{perspectives2024,buterin2023,lohn2022a}. Malicious actors
may for various reasons---including such unrelated to GPAI
models---become more or less numerous, able and willing to use GPAI
models. Over time, society could become better at managing risks from
models, meaning a model of a given level of capabilities may pose less
risk \citep{anderljung2023,bernardi2024,kapoor2024}. Under these
circumstances, compute thresholds may be adjusted upwards over time to
continue to focus on models posing a given level of risk.

Another development that can be relevant for adjusting compute
thresholds is computational price-performance improvements. In the
context of model training, computational price-performance refers to the
amount of compute that can be obtained for a given amount of money
\citep{hobbhahn2023}. Over time, the amount of training compute
available for a given amount of money increases due to innovations in
hardware manufacturing and design \citep{pilz2023a}. Importantly, such
improvements in computational price-performance do not influence the
amount of training compute necessary to achieve a given level of
capabilities, meaning they do not influence the level of risk from
models that cross a given compute threshold. However, improvements in
computational price-performance may increase regulatory burdens and
oversight costs: More actors may be able to afford the amount of compute
required to train models that meet the regulatory threshold
\citep{pilz2023a}. Again, this does not influence the level of risk
from models that cross a given compute threshold. But it might still be
an argument to adjust the compute threshold upwards to reduce regulatory
burdens and oversight costs. By doing so, however, regulators accept a
higher level of risk for society.\footnote{For a more comprehensive
  discussion of this issue, see \citet{pilz2023a}.}

The US AI EO tasks the Secretary of Commerce with updating its compute
thresholds, without specifying further details on how to do so
(Section~4.2(b)). We have provided what we believe to be the key
substantive considerations. The Secretary of Commerce should monitor
these developments and aim to increase understanding of risks from GPAI
models. On the procedural side, the US AI EO specifies that updates
should occur on a ``regular basis.'' An examination of whether a compute
threshold needs to be updated should happen at least every couple of
months, and maybe more frequently if the pace of algorithmic efficiency
or computational price-performance improvements increases.

The EU AI Act provides some guidance on how the European Commission
should update its thresholds for identifying GPAI models with systemic
risk, referring to ``technological developments, such as algorithmic
improvements or increased hardware efficiency'' (Article~51(3)). As
argued above, the mention of algorithmic (efficiency) improvements is
important. However, the reference to ``increased hardware efficiency''
is ambiguous and might reflect a misunderstanding. To the extent that
this term refers to innovations in hardware manufacturing and design
that may lead to computational price-performance improvements, as
explained above, corresponding developments do not influence the level
of risk from a given compute threshold. Instead, they only affect the
accessibility and proliferation of models that meet a given compute
threshold. As hardware becomes more efficient, the same amount of
compute becomes available at a lower cost and therefore potentially to a
broader range of actors. Consequently, hardware efficiency improvements
should be considered when adjusting compute thresholds, not because they
alter the relationship between training compute and model capabilities,
but because they may increase the number of actors in scope of a given
compute threshold. Furthermore, the EU AI Act guidance ignores two
crucial factors: better understanding of what constitutes risky models,
and societal developments that influence the level of risk posed by
models with a given level of capabilities.

\subsection{Alternative Metrics}\label{sec6.4:alternative-metrics}

Given the limitations of training compute, several other metrics may be
useful to distinguish between more and less risky models. For example,
the EU AI Act mentions a variety of metrics that should play a role in
identifying GPAI models with systemic risk: the number of model
parameters, the quantity and quality of training data, the amount of
training compute, input and output modalities and type, model
capabilities, a high impact on the internal market due to reach
(presumed if at least 10~000 registered business users in the Union),
and the number of registered end users (Annex XIII). Other metrics
commonly discussed include risk estimates, model capability evaluations,
effective compute, and some other metrics.

\subsubsection{Risk Estimates}\label{sec6.4.1:risk-estimates}

The most straightforward way to distinguish between more and less risky
models may be to directly measure, or rather estimate, the increase in
risk from a model. On the one hand, risk estimates are directly focused
on harm to individuals, groups, and society as a whole. Risk estimates
do not face the issue of potentially focusing on wrong proxies, such as
too little compute or harmless capabilities \citep{koessler2024}. On
the other hand, risk estimates are extremely hard to do well for an
emerging, general-purpose technology like GPAI. Little historical data
and a tremendous amount of risk scenarios mean risk estimates have to
rely on models and expert judgment, and they face the issue of missing
important risk scenarios \citep{schuett}. Overall, risk
estimates require a lot of effort to conduct, meaning they should not
replace compute thresholds as an initial filter. After that filter has
been applied, risk estimates are the ideal, but still highly immature,
metric to decide whether mitigation measures are necessary. Risk
thresholds defined in terms of risk estimates should thus not yet
determine, but only inform decisions about whether mitigation measures
are necessary \citep{koessler2024}.

\subsubsection{Model Capability
Evaluations}\label{sec6.4.2:model-capability-evaluations}

Another metric useful for distinguishing between more and less risky
models is model capability evaluations. Currently, the two most common
approaches to evaluations are ``benchmarking'' and ``red-teaming.''
Benchmarking involves using standardized tests to evaluate model
capabilities---these benchmarks allow comparisons between different
models. Well-known benchmarks include BBQ (bias; \citealt{parrish2022}),
GLUE (natural language understanding; \citealt{wang2019}), and MMLU
(general knowledge; \citealt{hendrycks2021}). Red-teaming involves less
standardized and more in-depth testing, often by domain experts, for
specific model capabilities and behaviors \citep{barrett2024,
jones2024}.

{Risk-tracking}: Model capabilities are attractive as a metric
  because they can be considered a proxy for risk and they are more
  closely related to risks than is training compute (\Cref{sec4:limitations-of-training-compute-relevant-for-gpai-regulation}).
  However, model capabilities are an imperfect proxy, too. Other
  important factors include the offense-defense balance of AI systems,
  the threat landscape, and societal vulnerability or adaptation.
  Regulating consistent levels of capability over time may therefore not
  be warranted. At the same time, model capabilities may create a false
  sense of exhaustiveness, potentially leading to underestimating the
  need for risk assessments \citep{jones2024}.

  \textbf{Difficult to measure}: Evaluating a model's capabilities is
  very challenging \citep{anwar2024,jones2024,reuel2024}. In
  particular, proving the absence of a capability is difficult, implying
  a potentially large amount of false negatives \citep{casper2024a}.
  Furthermore, models are highly prompt sensitive, meaning slight
  changes in inputs can lead to large differences in outputs
  \citep{mizrahi2024,ramesh2023,sclar2024}. The appropriate point
  in a model's lifecycle to evaluate capabilities can also lead to
  significant differences in results, as discussed in \Cref{sec2:training-compute}. In general,
  there still is large disagreement in the research community about what
  constitutes adequate model capability evaluations, as evidenced by many competing
  benchmarks for similar capabilities (e.g., MATH, MGSM, and GSM8K for
  mathematical capabilities) \citep{anwar2024,jones2024,
  reuel2024}. The field of dangerous model capability evaluations is
  especially nascent, with the first paper on the topic having been
  published about one year ago \citep{shevlane2023}. Finally,
  comprehensive model capability evaluations require significant effort \citep{anwar2024,jones2024,reuel2024}.

  \textbf{Easy to circumvent}: Model capability evaluations may be
  circumvented. Related to the previous point, easy-to-run evaluations
  may be easy to game, allowing companies to design models that perform
  poorly on benchmarks while still excelling at the capabilities the
  benchmarks are meant to measure. The lack of standardization makes it
  particularly easy to inflate evaluation results \citep{leech2024}.
  For example, companies may strategically choose which evaluations to
  conduct \citep{jones2024}. Companies or deceptive AI models could
  also strategically underperform on a given set of evaluations, a
  phenomenon referred to as ``sandbagging'' \citep{jarviniemi2024,
  vanderweij2024}.

  \textbf{Not measurable before development}: Training compute is
  usually known prior to development, whereas model capability
  evaluations can only be conducted during and after development. There
  have been attempts to forecast model capabilities
  \citep{phuong2024}; however, such forecasting is likely still highly
  unreliable, as indicated by high disagreement between forecasters.

  \textbf{Difficult to verify externally}: Model capabilities are also
  much harder to verify than training compute is. Since measuring model
  capabilities is more difficult, verifying these measurements is more
  difficult, too. Actors verifying model capabilities need more
  information and may even need to conduct model capability evaluations themselves.
  Regulators currently may not have the necessary expertise, and there
  is no mature third-party ecosystem, either. In the worst case,
  publishing model capability data might expose a company's algorithmic
  progress, potentially intensifying competitive dynamics in the AI
  industry.

  \textbf{Cost-tracking}: Model capabilities are highly correlated with
  the cost of developing the model, since increases in capabilities
  largely stem from increases in training compute (\Cref{sec3:features-of-training-compute-useful-for-gpai-regulation}).

Overall, training compute and model capabilities should be viewed as
complementary metrics. Training compute excels at providing a quick,
easily measurable, and externally verifiable metric to identify
potentially risky models and trigger regulatory oversight and further
scrutiny. Model capabilities are more expensive and harder to measure
but also provide more information about a model's risks. By using
compute thresholds as a first-pass filter and then applying capability
thresholds, regulators can create a more efficient and effective
regulatory process (\Cref{sec5.1:initial-filter}).

\subsubsection{Effective Compute}\label{sec6.4.3:effective-compute}

Algorithmic efficiency improvements, including through advancements in
model architecture and training methods, can over time reduce the amount
of compute required to train models to similar levels of performance
\citep{erdil2023,hernandez2020,ho2024,sherry2021}. As a result,
there is growing interest in using a metric called \emph{effective
compute}, which accounts for both increases in training compute and
algorithmic efficiency improvements. Effective compute describes the
equivalent increase in training compute that would be needed to match a
given model performance absent algorithmic efficiency improvements
\citep{ho2024}. Effective compute is a relative metric, as
improvements in algorithmic efficiency are measured against a specific
model performance. For example, effective compute may be expressed as
how many times more compute would have been required to train a model to
perform as well as a specific other model on a specific performance
metric, had there been no algorithmic improvements (e.g., $4\times$ effective
compute relative to GPT-4 test loss).

  \textbf{Risk-tracking}: Effective compute is as risk-tracking as the
  model performance metric it is tied to. For example, if this
  performance metric is a benchmark, then effective compute is a good of
  a proxy as the benchmark. If the performance metric is loss, then
  effective compute is as good as training compute. However, as with
  model capabilities, effective compute is only a proxy for risk.
  Regulating consistent levels of performance over time may therefore
  not be warranted. Depending on the performance metric, effective
  compute can even create a false sense of being a comprehensive measure
  of risk---as it combines training compute and model capabilities---potentially leading to underestimating the need for risk assessments.

\begin{figure}[ht!]
    \centering
    \centerline{\includegraphics[width=1.2\linewidth]{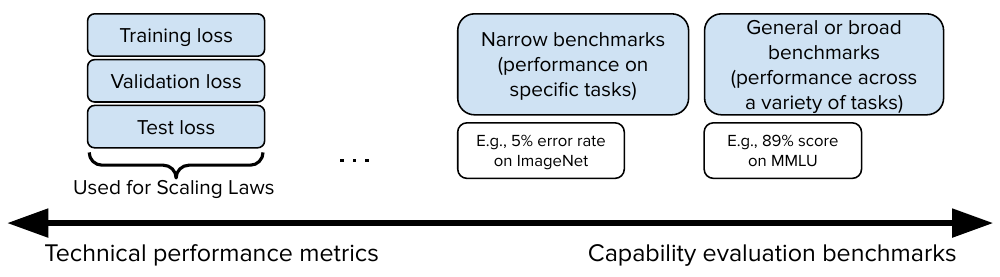}}
    \caption{Spectrum performance metrics effective compute can be tied to. Technical metrics like training loss, validation loss, and test loss scale more smoothly with compute, are easier to measure and verify, and are less context-dependent. In contrast, capability benchmarks may scale more suddenly with compute, are closer to real-world utility and impact, but can sometimes be more \emph{reductive} to specific tests.}
    \label{fig:fig9}
\end{figure}

  \textbf{Difficult to measure}: Effective compute is more difficult to
  measure than training compute is. Research on algorithmic efficiency
  improvements for AI models is sparse, with only a few papers published~\citep{erdil2023,hernandez2020,ho2024,sherry2021}. There is
  also no agreement on two key questions: which performance metric to
  use and at what point during the model lifecycle to assess performance
  on that metric. Any definition of effective compute needs to choose a
  performance metric (e.g., test loss) as the reference point that the
  amount of compute required is ``normalized'' to. Technical metrics
  like test loss are easier to measure and scale more smoothly with
  compute, but they may not accurately represent real-world impact,
  which makes them similar to training compute as a metric while
  concealing this fact. Metrics like model capability evaluations are
  more relevant for risk but harder to measure and subject to sudden or
  unpredictable increases in performance, such as the emergence of new
  capabilities (\Cref{fig:fig9}).
  The appropriate point in the model's lifecycle to measure performance
  on a given metric also remains unclear. As discussed in \Cref{sec2:training-compute}, performance
  gains can result from post-training enhancements like fine-tuning or
  prompting methods, often resulting in relatively larger increases in
  effective compute, primarily driven by the performance improvements
  achieved with a comparatively small additional compute investment.
  Furthermore, there is ambiguity regarding which prompting techniques
  (such zero-shot, few-shot, or chain-of-thought) should be allowed when
  measuring model capabilities, adding to the difficulty of making
  reliable statements about effective compute.

  \textbf{Easy to circumvent}: The ease of circumventing effective
  compute depends on the chosen performance metric. Quantitative metrics
  like training loss or validation performance are more robust, but if
  based on evaluations or benchmarks, effective compute faces similar
  circumvention risks as those of model capability evaluations (\Cref{sec6.4.2:model-capability-evaluations}).
  Companies could game evaluations by designing models to underperform
  while still excelling at intended capabilities, strategically
  selecting evaluations, or intentionally underperforming
  (``sandbagging'').

  \textbf{Not measurable before development}: Training compute is
  usually known prior to development, whereas effective compute requires
  analysis of performance after development. As with model capabilities,
  performance can be forecasted, but such forecasts are still highly
  unreliable.

  \textbf{Difficult to verify externally}: Verifying effective compute
  may require access to more proprietary information than does training
  compute. Publishing effective compute data alongside standard training
  compute data could expose a company's algorithmic progress,
  potentially intensifying competitive dynamics in the AI industry.

  \textbf{Cost-tracking}: Effective compute is correlated with
  the cost of developing the model, since increases in performance
  largely stem from increases in training compute (\Cref{sec3:features-of-training-compute-useful-for-gpai-regulation}).

In conclusion, there currently are practical challenges to using
effective compute as a metric. More research is needed to develop
standardized ways to measure effective compute that address these
challenges. If successful, an effective compute measure could become the
``GDP per capita'' of AI performance (e.g., by creating ``a nominal
basket of benchmarks''). Effective compute is currently most suitable
for internal company use, where the necessary tools and insights for
accurate assessment are readily available. For example, companies can
use effective compute for selecting checkpoints for their safety
policies \citep{anthropic2023,dragan2024,openai2023a}. External
entities, including regulatory bodies, may struggle to accurately assess
effective compute due to the lack of standardized methods and limited
access to models. Moreover, regulatory bodies may not yet possess the
technical expertise required to accurately assess effective compute.
Given these challenges, we currently recommend not using effective
compute for regulatory purposes.

\subsubsection{Parameters, Data, and Other Metrics}\label{sec6.4.4:parameters-data-and-other-metrics}

This section provides an initial analysis and comparison of other
metrics to the more extensively discussed ones above. The metrics in
this section include most of the main variables in model development and
deployment. In the discussion of each metric, we assume the values of
all other variables remain fixed (e.g., when talking about the
correlation between model architecture and risk we assume a fixed amount
of training compute). We highlight that the discussion in this section
is preliminary and many points we make need further work and empirical
investigation.

  \emph{Model architecture and training algorithm}: The model's
  architecture and training algorithm influence its capabilities and
  thus risk, but empirically to a much lesser extent than the amount of
  training compute \citep{ho2024,kaplan2020,sutton2019}. Moreover,
  these metrics are hard to quantify, they often involve step changes,
  and they are multidimensional with complex synergies between different
  dimensions (e.g., a certain number of layers might work well with a
  certain learning rate but not another).

  \emph{Number of model parameters}: The number of model parameters
  tracks capabilities and thus risk \citep{kaplan2020} and can simply
  be counted \citep{villalobos2022}. However, certain model
  architectures like Mixture of Experts (MoE) can significantly change
  the number of model parameters without necessarily increasing the
  model's capabilities and risk \citep{villalobos2022}, and the number
  of model parameters can be manipulated through post-training
  techniques like model pruning without necessarily decreasing the
  model's capabilities and risk \citep{cheng2023}. What is more, the
  number of model parameters affects the necessary amount of training
  compute and is thus captured by that metric. More training compute is
  required to handle more parameters trained on a given amount of data.
  Additionally, developers can be expected to aim for an optimal ratio
  between the number of model parameters and the amount of training data
  as described by scaling laws. Therefore, for a given architecture and
  assuming training compute allocation was optimal according to scaling
  laws, training compute implies the number of model parameters.

  \emph{Amount and quality of training data}: The amount of training
  data affects model capabilities and risk \citep{kaplan2020} and can
  be quantified in tokens or bytes \citep{villalobos2024}. However,
  data quality is also crucial, influencing capabilities across
  dimensions like information density and diversity, yet lacking
  objective or standardized measurement methods \citep{mitchell2023}.
  Importantly, amount and quality are interrelated - more data does not
  guarantee better performance with poor quality, while high-quality
  data can compensate for smaller sizes \citep{evans2024,
  gunasekar2023}. Additionally, the number of epochs or passes over
  the training data during the training process also affects the total
  compute requirements, even with a fixed dataset size. Developers
  optimize the ratio of parameters, data size, data quality, and number
  of epochs per scaling laws. Thus, for a given architecture with
  optimal compute allocation, training compute serves as an indirect
  proxy for the amount of training data utilized, as well as the number
  of training epochs but not for the quality.

  \emph{Number of users}: The number of users correlates with risks from
  accidents and, to some extent, misuse (the larger the user base, the
  more likely misuse occurs). However, it may not correlate with risks
  from misalignment or risks that arise before deployment, such as
  malicious actors stealing and misusing the model. This metric is
  difficult to estimate before deployment and even more so before
  development.

  \emph{Applications}: While applications are often known before
  deployment (except if model weights are publicly released), many GPAI
  models are used in general-purpose tools, like GPT-4 powering ChatGPT.
  Before development, it often does not make sense to consider
  applications, as GPAI models, by definition, have broad capabilities
  and can be used for many downstream applications \citep{jones2023}.

  \emph{Harm}: This metric is the materialization of risk. However, it
  is not known in advance of development or deployment. The best we can
  do in advance of development and deployment is to estimate the
  likelihood and severity of harm, that is, estimate risk (\Cref{sec6.4.1:risk-estimates}). Of
  course, such risk estimates should be updated with the data gathered
  through monitoring, and development or deployment should be
  discontinued at any time if a particular number or type of incidents
  occur. However, at that point, the harm from those incidents cannot be
  undone. Therefore, in particular for irreversible and large-scale
  harm, in addition to ex post metrics that measure risk in hindsight,
  we need ex ante metrics that measure risk in advance.

\begin{table}[ht!]
    \centering
    \begin{adjustwidth}{-0.2\linewidth}{-0.2\linewidth}
    \resizebox{\linewidth}{!}{%
    \begin{tikzpicture}
    \tikzstyle{cell} = [minimum width=2.75cm, minimum height=1.7cm,text width=2.75cm, align=center,myblack]
    \tikzstyle{L} = [cell, fill=myred!80]
    \tikzstyle{LM} = [cell, fill=myred!60]
    \tikzstyle{M} = [cell, fill=myorange!80]
    \tikzstyle{MH} = [cell, fill=myorange!60]
    \tikzstyle{H} = [cell, fill=mygreen!80]
    
    \node[cell,fill=background] (A1) {Features/ \\Metrics};
    \node[right =0.5mm of A1,cell,fill=background] (B1) {\textbf{Correlation with risk}};
    \node[right =0.5mm of B1,cell,fill=background] (C1) {\textbf{Ease of measurement}};
    \node[right =0.5mm of C1,cell,fill=background] (D1) {\textbf{Difficulty of circumvention}};
    \node[right =0.5mm of D1,cell,fill=background] (E1) {\textbf{Measurable before development and deployment}};
    \node[right =0.5mm of E1,cell,fill=background] (F1) {\textbf{Externally verifiable}};
    \node[right =0.5mm of F1,cell,fill=background] (G1) {\textbf{Correlation with cost}};
    \node[right =0.5mm of G1,cell,fill=background] (H1) {\textbf{Frequency of updates required}};

    \node[below =0.5mm of A1,cell,fill=background] (A2) {\textbf{Model architecture and training algorithm}};
    \node[below =0.5mm of B1,cell,fill=background,L] (B2) {Low};
    \node[below =0.5mm of C1,cell,fill=background,L] (C2) {Low};
    \node[below =0.5mm of D1,cell,fill=background,LM] (D2) {Low to medium};
    \node[below =0.5mm of E1,cell,fill=background,H] (E2) {Yes};
    \node[below =0.5mm of F1,cell,fill=background,M] (F2) {Medium};
    \node[below =0.5mm of G1,cell,fill=background,LM] (G2) {Low to medium};
    \node[below =0.5mm of H1,cell,fill=background,M] (H2) {Medium};

    \node[below =0.5mm of A2,cell,fill=background] (A3) {\textbf{Number of model parameters}};
    \node[below =0.5mm of B2,cell,fill=background,LM] (B3) {Low to medium};
    \node[below =0.5mm of C2,cell,fill=background,H] (C3) {High};
    \node[below =0.5mm of D2,cell,fill=background,M] (D3) {Medium};
    \node[below =0.5mm of E2,cell,fill=background,H] (E3) {Yes};
    \node[below =0.5mm of F2,cell,fill=background,H] (F3) {High};
    \node[below =0.5mm of G2,cell,fill=background,MH] (G3) {Medium to high};
    \node[below =0.5mm of H2,cell,fill=background,M] (H3) {Medium};    

    \node[below =0.5mm of A3,cell,fill=background] (A4) {\textbf{Amount and quality of training data}};
    \node[below =0.5mm of B3,cell,fill=background,M] (B4) {Medium};
    \node[below =0.5mm of C3,cell,fill=background,M] (C4) {Medium};
    \node[below =0.5mm of D3,cell,fill=background,MH] (D4) {Medium to high};
    \node[below =0.5mm of E3,cell,fill=background,H] (E4) {Yes};
    \node[below =0.5mm of F3,cell,fill=background,MH] (F4) {Medium to high};
    \node[below =0.5mm of G3,cell,fill=background,MH] (G4) {Medium to high};
    \node[below =0.5mm of H3,cell,fill=background,M] (H4) {Medium}; 

    \node[below =0.5mm of A4,cell,fill=background] (A5) {\textbf{Training compute}};
    \node[below =0.5mm of B4,cell,fill=background,M] (B5) {Medium};
    \node[below =0.5mm of C4,cell,fill=background,H] (C5) {High};
    \node[below =0.5mm of D4,cell,fill=background,H] (D5) {High};
    \node[below =0.5mm of E4,cell,fill=background,H] (E5) {Yes};
    \node[below =0.5mm of F4,cell,fill=background,H] (F5) {High};
    \node[below =0.5mm of G4,cell,fill=background,H] (G5) {High};
    \node[below =0.5mm of H4,cell,fill=background,M] (H5) {Medium};    

    \node[below =0.5mm of A5,cell,fill=background] (A6) {\textbf{Loss}};
    \node[below =0.5mm of B5,cell,fill=background,MH] (B6) {Medium to high};
    \node[below =0.5mm of C5,cell,fill=background,MH] (C6) {Medium to high};
    \node[below =0.5mm of D5,cell,fill=background,H] (D6) {High};
    \node[below =0.5mm of E5,cell,fill=background,M] (E6) {Yes before deployment; no before development};
    \node[below =0.5mm of F5,cell,fill=background,M] (F6) {Medium};
    \node[below =0.5mm of G5,cell,fill=background,MH] (G6) {Medium to high};
    \node[below =0.5mm of H5,cell,fill=background,M] (H6) {Medium};  

    \node[below =0.5mm of A5,cell,fill=background] (A6) {\textbf{Loss}};
    \node[below =0.5mm of B5,cell,fill=background,MH] (B6) {Medium to high};
    \node[below =0.5mm of C5,cell,fill=background,MH] (C6) {Medium to high};
    \node[below =0.5mm of D5,cell,fill=background,H] (D6) {High};
    \node[below =0.5mm of E5,cell,fill=background,M] (E6) {Yes before deployment; no before development};
    \node[below =0.5mm of F5,cell,fill=background,M] (F6) {Medium};
    \node[below =0.5mm of G5,cell,fill=background,MH] (G6) {Medium to high};
    \node[below =0.5mm of H5,cell,fill=background,M] (H6) {Medium};    

    \node[below =0.5mm of A6,cell,fill=background] (A7) {\textbf{Model capability evaluations}};
    \node[below =0.5mm of B6,cell,fill=background,MH] (B7) {Medium to high};
    \node[below =0.5mm of C6,cell,fill=background,M] (C7) {Medium};
    \node[below =0.5mm of D6,cell,fill=background,LM] (D7) {Low to medium};
    \node[below =0.5mm of E6,cell,fill=background,M] (E7) {Yes before deployment; no before development};
    \node[below =0.5mm of F6,cell,fill=background,M] (F7) {Medium};
    \node[below =0.5mm of G6,cell,fill=background,MH] (G7) {Medium to high};
    \node[below =0.5mm of H6,cell,fill=background,M] (H7) {Medium};  

    \node[below =0.5mm of A7,cell,fill=background] (A8) {\textbf{Effective compute}};
    \node[below =0.5mm of B7,cell,fill=background,MH] (B8) {Medium to high};
    \node[below =0.5mm of C7,cell,fill=background,M] (C8) {Medium};
    \node[below =0.5mm of D7,cell,fill=background,M] (D8) {Medium};
    \node[below =0.5mm of E7,cell,fill=background,M] (E8) {Yes before deployment; no before development};
    \node[below =0.5mm of F7,cell,fill=background,M] (F8) {Medium};
    \node[below =0.5mm of G7,cell,fill=background,MH] (G8) {Medium to high};
    \node[below =0.5mm of H7,cell,fill=background,MH] (H8) {Low to medium};

    \node[below =0.5mm of A8,cell,fill=background] (A9) {\textbf{Number of users}};
    \node[below =0.5mm of B8,cell,fill=background,M] (B9) {High for some risks; low for other risks};
    \node[below =0.5mm of C8,cell,fill=background,H] (C9) {High};
    \node[below =0.5mm of D8,cell,fill=background,H] (D9) {High};
    \node[below =0.5mm of E8,cell,fill=background,M] (E9) {Sometimes before deployment; no before development};
    \node[below =0.5mm of F8,cell,fill=background,M] (F9) {Medium};
    \node[below =0.5mm of G8,cell,fill=background,M] (G9) {Medium};
    \node[below =0.5mm of H8,cell,fill=background,L] (H9) {High};  

    \node[below =0.5mm of A9,cell,fill=background] (A10) {\textbf{Applications}};
    \node[below =0.5mm of B9,cell,fill=background,M] (B10) {High for some risks; low for other risks};
    \node[below =0.5mm of C9,cell,fill=background,M] (C10) {Medium};
    \node[below =0.5mm of D9,cell,fill=background,M] (D10) {Medium};
    \node[below =0.5mm of E9,cell,fill=background,M] (E10) {Sometimes};
    \node[below =0.5mm of F9,cell,fill=background,M] (F10) {Medium};
    \node[below =0.5mm of G9,cell,fill=background,M] (G10) {Medium};
    \node[below =0.5mm of H9,cell,fill=background,H] (H10) {Low};   

    \node[below =0.5mm of A10,cell,fill=background] (A11) {\textbf{Risk estimates}};
    \node[below =0.5mm of B10,cell,fill=background,H] (B11) {High};
    \node[below =0.5mm of C10,cell,fill=background,L] (C11) {Low};
    \node[below =0.5mm of D10,cell,fill=background,M] (D11) {Medium};
    \node[below =0.5mm of E10,cell,fill=background,H] (E11) {Yes};
    \node[below =0.5mm of F10,cell,fill=background,M] (F11) {Medium};
    \node[below =0.5mm of G10,cell,fill=background,MH] (G11) {Medium to high};
    \node[below =0.5mm of H10,cell,fill=background,H] (H11) {Low};  

    \node[below =0.5mm of A11,cell,fill=background] (A12) {\textbf{Harm}};
    \node[below =0.5mm of B11,cell,fill=background,H] (B12) {High};
    \node[below =0.5mm of C11,cell,fill=background,L] (C12) {Low};
    \node[below =0.5mm of D11,cell,fill=background,M] (D12) {Medium};
    \node[below =0.5mm of E11,cell,fill=background,L] (E12) {No};
    \node[below =0.5mm of F11,cell,fill=background,M] (F12) {Medium};
    \node[below =0.5mm of G11,cell,fill=background,MH] (G12) {Medium to high};
    \node[below =0.5mm of H11,cell,fill=background,H] (H12) {Low};          
    \end{tikzpicture}%
    }
    \end{adjustwidth}
    \caption{First-pass ranking of potential metrics for GPAI
regulation regarding the features discussed in \Cref{sec3:features-of-training-compute-useful-for-gpai-regulation}.}
    \label{tab:1}
\end{table}

Complementing mere compute thresholds with other metrics becomes
relevant to the extent that scaling laws cease to hold and training
compute becomes a worse proxy for risk
(\Cref{sec4:limitations-of-training-compute-relevant-for-gpai-regulation}).
Particularly relevant combinations may include training compute and
model capability evaluations (to ensure catching the most capable
models) and training compute and number of users (to ensure catching the
most widely used models). However, any threshold that is supposed to
serve as an initial filter to identify models of potential concern
should be based on metrics that can be measured easily and early in the
model lifecycle. This, at least currently, excludes model capability
evaluations, and at least before deployment, the number of users.

\section{Conclusion}\label{sec7:conclusion}

While not perfect, compute thresholds are currently one of the key tools
in GPAI regulation. They offer a risk-correlated, easily measurable, and
externally verifiable metric that can inform regulatory decisions while
minimizing circumvention and targeting the most well-resourced actors.
They are currently the best tool for identifying potentially risky
models and triggering regulatory oversight and further scrutiny.

Further research is necessary on many questions related to compute
thresholds. In particular, the trends relevant for updating compute
thresholds require further research---both empirical research on their
current development and theoretical research on their future development
(\Cref{sec6.3:why-when-and-how-to-update-compute-thresholds}).
Moreover, the usefulness of metrics other than training compute should
be studied further, especially effective compute, which we consider a
highly promising metric (\Cref{sec6.4.3:effective-compute}), and how
to combine training compute with other metrics such as model capability
evaluations or the number of users (\Cref{sec6.4.4:parameters-data-and-other-metrics}).
Another area of research we would like to highlight concerns the
enforcement of compute thresholds. What information do regulators need?
What process should regulators rely on to ensure they get that
information (e.g., should they inspect companies)? What special
considerations apply if companies are not physically located within the
territory of the regulator's jurisdiction?

Finally, compute thresholds are not sufficient for GPAI regulation---they are one tool among many. Their effectiveness depends on the
specific context and the design of the overall regulatory framework. In
particular, the discussion of whether capability thresholds are better
than compute thresholds is misguided. Capability thresholds are
supplemental, and while they might play a more important role in the
future, compute thresholds are likely to remain a valuable tool in GPAI
regulation, too. By understanding the strengths and limitations of
compute thresholds, policymakers can make informed decisions about when
and how to use them as part of a comprehensive approach to GPAI
regulation.

\section*{Acknowledgments}

We are grateful for valuable input from the following individuals (in
alphabetical order by surname): Neel Alex, Markus Anderljung, Mauricio
Baker, Lukas Berglund, Benjamin Bucknall, Alan Chan, Stephen Clare,
Connor Dunlop, Jimmy Farrell, James Gealy, John Halstead, Marius
Hobbhahn, Nicolas Moës, James Petrie, Konstantin Pilz, Jonas Sandbrink, Risto Uuk, and many others. All
views and remaining errors are our own.

\clearpage

\addcontentsline{toc}{section}{References}
\bibliography{references}
\bibliographystyle{icml2024}

\end{document}